\documentclass[twocolumn]{aastex631}

\usepackage{booktabs} 
\usepackage{hyperref}

\begin{document}

\title{The Complete Catalog of Gamma-Ray Transients Observed by GRBAlpha \& VZLUSAT-2 CubeSat Missions}

\correspondingauthor{M. Dafčíková}
\email{mdafcikova@mail.muni.cz}

\author[0009-0000-0365-5940]{Marianna Dafčíková}
\affiliation{Department of Theoretical Physics and Astrophysics, Faculty of Science, Masaryk University, Kotlářská 267/2, Brno 611 37, Czech Republic}

\author[0000-0003-3994-7528]{Jakub Řípa}
\affiliation{Department of Theoretical Physics and Astrophysics, Faculty of Science, Masaryk University, Kotlářská 267/2, Brno 611 37, Czech Republic}

\author[0000-0001-5449-2467]{András Pál}
\affiliation{Konkoly Observatory, Research Centre for Astronomy and Earth Sciences, Budapest, Hungary}

\author[0000-0003-0392-0120]{Norbert Werner}
\affiliation{Department of Theoretical Physics and Astrophysics, Faculty of Science, Masaryk University, Kotlářská 267/2, Brno 611 37, Czech Republic}

\author[0009-0008-2122-9097]{Michaela Ďuríšková}
\affiliation{Department of Theoretical Physics and Astrophysics, Faculty of Science, Masaryk University, Kotlářská 267/2, Brno 611 37, Czech Republic}

\author[0000-0002-0921-8837]{Yasushi Fukazawa}
\affiliation{Department of Physics, Graduate School of Advanced Science and Engineering, Hiroshima University, Higashi-Hiroshima, Japan}

\author{Martin Kolář}
\affiliation{Department of Theoretical Physics and Astrophysics, Faculty of Science, Masaryk University, Kotlářská 267/2, Brno 611 37, Czech Republic}

\author{László Mészáros}
\affiliation{Konkoly Observatory, Research Centre for Astronomy and Earth Sciences, Budapest, Hungary}

\author[0000-0003-0975-9753]{Filip Münz}
\affiliation{Department of Theoretical Physics and Astrophysics, Faculty of Science, Masaryk University, Kotlářská 267/2, Brno 611 37, Czech Republic}

\author{Masanori Ohno}
\affiliation{Department of Physics, Graduate School of Advanced Science and Engineering, Hiroshima University, Higashi-Hiroshima, Japan}

\author{Lea Szakszonová}
\affiliation{Department of Theoretical Physics and Astrophysics, Faculty of Science, Masaryk University, Kotlářská 267/2, Brno 611 37, Czech Republic}

\author[0000-0001-6314-5897]{Hiromitsu Takahashi}
\affiliation{Department of Physics, Graduate School of Advanced Science and Engineering, Hiroshima University, Higashi-Hiroshima, Japan}

\author{Masato Yokota}
\affiliation{Department of Physics, Graduate School of Advanced Science and Engineering, Hiroshima University, Higashi-Hiroshima, Japan}

\author[0000-0001-6131-4802]{Jean-Paul Breuer}
\affiliation{Department of Physics, Graduate School of Advanced Science and Engineering, Hiroshima University, Higashi-Hiroshima, Japan}

\author[0000-0002-5617-3117]{Hsiang-Kuang Chang}
\affiliation{Institute of Astronomy, National Tsing Hua University, Hsinchu, Taiwan, Republic of China}

\author{Balázs Csák}
\affiliation{Konkoly Observatory, Research Centre for Astronomy and Earth Sciences, Budapest, Hungary}

\author[0000-0002-3704-6454]{Vladimír Dániel}
\affiliation{VZLU AEROSPACE, a.s., Prague, Czech Republic}

\author{Juraj Dudáš}
\affiliation{VZLU AEROSPACE, a.s., Prague, Czech Republic}

\author{Marcel Frajt}
\affiliation{Spacemanic Ltd., Bratislava, Slovakia}

\author[0009-0009-7079-5981]{Gábor Galgóczi}
\affiliation{Wigner Research Centre for Physics, Budapest, Hungary}

\author[0000-0003-3440-855X]{Peter Hanák}
\affiliation{Faculty of Aeronautics, Technical University of Kosice, Košice, Slovakia}

\author{Filip Hroch}
\affiliation{Department of Theoretical Physics and Astrophysics, Faculty of Science, Masaryk University, Kotlářská 267/2, Brno 611 37, Czech Republic}

\author[0000-0001-8551-2002]{Chin-Ping Hu}
\affiliation{Department of Physics, National Changhua University of Education, Changhua City, Taiwan, Republic of China}

\author{Jan Hudec}
\affiliation{Spacemanic Ltd., Bratislava, Slovakia}

\author{Nikola Husáriková}
\affiliation{Department of Theoretical Physics and Astrophysics, Faculty of Science, Masaryk University, Kotlářská 267/2, Brno 611 37, Czech Republic}

\author{Yuto Ichinohe}
\affiliation{RIKEN Nishina Center for Accelerator-Based Science, Saitama, Japan}

\author{Jakub Kapuš}
\affiliation{Spacemanic Ltd., Bratislava, Slovakia}

\author{Miroslav Kasal}
\affiliation{Department of Radio Electronics, Faculty of Electrical Engineering and Communication, Brno University of Technology, Brno, Czech Republic}

\author{Martin Koleda}
\affiliation{Needronix Ltd., Bratislava, Slovakia}

\author{Róbert László}
\affiliation{Needronix Ltd., Bratislava, Slovakia}

\author[0000-0002-8578-7775]{Chih-Hsun Lin}
\affiliation{Institute of Physics, Academia Sinica, Taipei, Taiwan, Republic of China}

\author{Tsung-Che Liu}
\affiliation{Department of Applied Physics, National Pingtung University, Pingtung, Taiwan, Republic of China}

\author[0000-0001-7263-0296]{Tsunefumi Mizuno}
\affiliation{Department of Physics, Graduate School of Advanced Science and Engineering, Hiroshima University, Higashi-Hiroshima, Japan}

\author[0000-0003-2930-350X]{Kazuhiro Nakazawa}
\affiliation{Department of Physics, Nagoya University, Nagoya, Aichi, Japan}

\author{Hirokazu Odaka}
\affiliation{Department of Physics, The University of Tokyo, Bunkyo-ku, Tokyo, Japan}

\author[0000-0002-6972-2402]{Michal Pazderka}
\affiliation{Department of Theoretical Physics and Astrophysics, Faculty of Science, Masaryk University, Kotlářská 267/2, Brno 611 37, Czech Republic}
\affiliation{R\&D Center for Low-Cost Plasma and Nanotechnology Surface Modifications, CEPLANT, Department of Plasma Physics and Technology, Faculty of Science, Brno, Czech Republic}

\author[0000-0001-7693-9901]{Aleš Povalač}
\affiliation{Department of Radio Electronics, Faculty of Electrical Engineering and Communication, Brno University of Technology, Brno, Czech Republic}

\author{Maksim Rezenov}
\affiliation{Spacemanic Ltd., Bratislava, Slovakia}

\author{Martin Sabol}
\affiliation{VZLU AEROSPACE, a.s., Prague, Czech Republic}

\author[0009-0009-6295-3142]{Kaustubha Sen}
\affiliation{Institute of Astronomy, National Tsing Hua University, Hsinchu, Taiwan, Republic of China}

\author[0000-0002-0314-3651]{Miroslav Šmelko}
\affiliation{EDIS vvd., Košice, Slovakia}

\author[0000-0002-2768-4030]{Petr Svoboda}
\affiliation{VZLU AEROSPACE, a.s., Prague, Czech Republic}

\author{Martin Topinka}
\affiliation{INAF - Osservatorio Astronomico di Cagliari, Selargius (CA), Italy}
\affiliation{Charles University, Faculty of Mathematics and Physics, Astronomical Institute, Prague, Czech Republic}

\author[0009-0003-7940-8016]{Che-Chih Tsao}
\affiliation{Department of Power Mechanical Engineering, National Tsing Hua University, Hsinchu, Taiwan, Republic of China}

\author[0000-0003-1512-4287]{Tomáš Urbanec}
\affiliation{Department of Radio Electronics, Faculty of Electrical Engineering and Communication, Brno University of Technology, Brno, Czech Republic}

\author[0000-0003-1514-3242]{Ivo Veřtát}
\affiliation{University of West Bohemia, Department of Applied Electronics and Telecommunications, Plzeň, Czech Republic}

\author[0009-0006-3048-4335]{Tomáš Vítek}
\affiliation{Department of Theoretical Physics and Astrophysics, Faculty of Science, Masaryk University, Kotlářská 267/2, Brno 611 37, Czech Republic}

\author[0009-0007-1586-3940]{Chih-En Wu}
\affiliation{Institute of Astronomy, National Tsing Hua University, Hsinchu, Taiwan, Republic of China}

\begin{abstract}

We present the largest sample of gamma-ray transients observed by any CubeSat mission so far. Observations were acquired by a 1U CubeSat GRBAlpha, the smallest astrophysical space observatory, and a 3U CubeSat VZLUSAT-2. Both missions were technological pathfinders and carried a novel CsI scintillator-based detector read-out by silicon photomultipliers. 
They operated on Sun-synchronous low Earth orbits below 550 km for about four years; GRBAlpha between March 2021 and June 2025 while VZLUSAT-2 between January 2022 and November 2025.
Despite being technological experiments, they observed over 300 gamma-ray transients including gamma-ray bursts (GRBs), solar flares, soft gamma repeaters and one outburst from an X-ray binary. Among these are the two brightest GRBs ever observed, GRB 221009A and GRB 230307A, without saturation and GRBs at redshifts up to $z=4.2$. 
GRBAlpha also contributed to the InterPlanetary Network. 
Regular monitoring of transients was demonstrated by a detection rate of two transients or one GRB a week and the shortest time between two subsequent detections of only 42 minutes. 
We show that a constellation of nanosatellites around the Earth would observe at least $60~\%$ of Fermi/GBM GRBs with $5\sigma$ significance and over $90~\%$ at $3\sigma$ level. 
GRBAlpha and VZLUSAT-2 prove that routine monitoring of the gamma-ray sky can also be done by low-cost and quickly developed nanosatellite missions.

\end{abstract}


\section{Introduction} \label{sec:intro}

The gamma-ray sky, dominated by the cosmic X-ray background and emission from within our own Galaxy \citep{Ackermann2012,Galgoczi2021a}, can temporarily get completely outshined by various short-lived events. 
Gamma-ray bursts (GRBs) are the most energetic not only among other gamma-ray transients but in the entire universe \citep{Frail2001,Meszaros2006}. 
The widely accepted fireball model \citep{fireball_1,fireball_2} describes the GRB prompt emission coming from shocks inside a jet powered by a ``hidden'' central engine and the GRB afterglow as a result of the jet interacting with an ambient medium. 
The central engine will turn on either by a~collapse of a~massive star producing \textit{long} GRB lasting $\sim10$~s, or by merging compact objects resulting in \textit{short} GRB up to $\sim2$~s long. 
The first evidence for the origin of long GRBs came nearly three decades ago from the spatially and temporally coincident GRB 980425 and SN 1998bw \citep{Galama1998}, and was definitely confirmed few years later by spectroscopic observations of a~supernova SN 2003dh associated with GRB 030329 \citep{GRB030329_SN2003dh}.
The connection of at least some short GRBs to merging compact objects has been confirmed more recently by, to date, the only direct observation of gamma-rays and gravitational waves (GWs) from the same source in the case of GRB 170817A and GW 170817 \citep{Abbott2017,Goldstein2017}.
Despite the progress achieved in the past two decades,
the fundamentals, such as details of the central engine, emission mechanisms or jet structure, are yet to be fully understood.  
Moreover, new challenges appeared such as two long GRBs, GRB 211211A \citep{Rastinejad2022,Troja2022} and GRB 230307A \citep{Levan2023,Dichiara2023}, possibly accompanying neutron star mergers.
GRBs are key elements in searches for electromagnetic counterparts to other messengers, especially GWs \citep{Burns2020,Margutti2021} and neutrinos \citep{Waxman1997,Kimura2017}. 
As the O4 LIGO/Virgo/KAGRA (LVK) run \citep{2025arXiv250818082T,2026arXiv260527225T} did not bring any certain joint GW and GRB observations, the perspective for future remains challenging.
To increase the probability of future joint detections, a~full sky coverage by gamma-ray monitors at all times is essential.

Efforts have been made to provide rapid follow-up observations of GRBs across the electromagnetic spectrum such as The Neil Gehrels Swift Observatory \citep{Gehrels2004}. Upon receiving a gamma-ray trigger, it slews toward the calculated position and within minutes begins follow-up measurements in optical to X-ray.
However, their narrow instantaneous field of view limits the detection rate.
Furthermore, on near-Earth orbits, large fraction of the sky is blocked by the Earth and thus a~single mission is unable to provide the~full sky coverage crucial for breakthrough discoveries in multi-messenger astrophysics.

To provide a quick localization of gamma-ray events across the full sky at a reasonable cost and on a~short timeline, large constellations of small satellites have been proposed. 
Such a~fleet could rapidly localize GRBs via the triangulation technique \citep{Hurley2020,Ohno2020,Sanna2021,Thomas2023} and navigate many other observatories, both on ground and in space, toward the source.
Numerous pathfinders have been launched in recent years including 
BurstCube \citep{burstcube}, EIRSAT-1 \citep{eirsat1}, GRBAlpha \citep{Pal2021,Pal2023}, GRBBeta \citep{grbbeta,Ripa26_grbbeta}, GRID \citep{GRID}, 
HERMES \citep{HERMES}, SpIRIT \citep{SpIRIT}, or VZLUSAT-2 \citep{Daniel2020}.

One of the proposed constellations is the CAMELOT mission \citep{CAMELOT}. Its first technological pathfinder, the GRBAlpha 1U CubeSat, successfully completed its more than four years of operations in June 2025. The lessons learned from this mission were implemented into its successor; the GRBBeta 2U CubeSat launched in July 2024 and currently still operational. Moreover, the detector system was also employed in the Czech VZLUSAT-2 nanosatellite mission which operated until November 2025. 
In this paper, we summarize all of the astrophysical transients observed by the GRBAlpha and VZLUSAT-2 missions and discuss the prospects for a future scientific constellation of gamma-ray monitors important for multi-messenger astrophysics. 
The data products are publicly available and can be freely used for analysis of transients of interest, as well as for cross-correlation and comparative studies with other gamma-ray missions.

The paper is structured as follows. Section~\ref{sec:missions} introduces the GRBAlpha and VZLUSAT-2 missions and their instruments. Section~\ref{sec:methods} describes the observations and methods used in our analysis. The catalog is presented in section~\ref{sec:catalogue} and in section~\ref{sec:fermi-gbm} we compare the observed GRBs to Fermi/GBM measurements. 
Finally, section~\ref{sec:discussion_summary} provides a summary of this work and discusses the main results.

\section{Missions} \label{sec:missions}

GRBAlpha \citep{Pal2021,Pal2023} was the first technological pathfinder toward the CAMELOT constellation of nanosatellites to monitor and rapidly localize GRBs across the entire sky. On a 1U CubeSat platform it carried a gamma-ray detector based on a CsI(Tl) scintillator read-out by silicon photomultipliers (SiPMs). The detector is sensitive from about 40~keV to nearly 1~MeV with the maximum effective area of 54~cm$^2$ around 100~keV. The satellite was launched on 22 March 2021 to a Sun-synchronous orbit (SSO) with an inclination of 97.5 degrees and an altitude of approximately 550~km. GRBAlpha deorbited on 9 June 2025 \citep{GRBAlpha_reentry} after over 1500 
days of successful operations.

VZLUSAT-2 \citep{Daniel2020} was a 3U CubeSat launched on 13 January 2022 to an orbit nearly identical to that of GRBAlpha but at a slightly lower altitude of about 535~km. It re-entered the atmosphere on 30 November 2025. The primary instruments were optical cameras dedicated to Earth observations. Among the secondary payloads were various X-ray/gamma-ray detectors including two of those developed for GRBAlpha. 
On VZLUSAT-2, they were placed under the solar panels so the effective area is slightly decreased in low energies.
This work reports only on the results obtained from the CsI scintillator-based detectors.

\section{Methodology} \label{sec:methods}

\subsection{Data and Observations} 
\label{sec:energy}

\begin{figure}
    \centering
    \includegraphics[width=\linewidth]{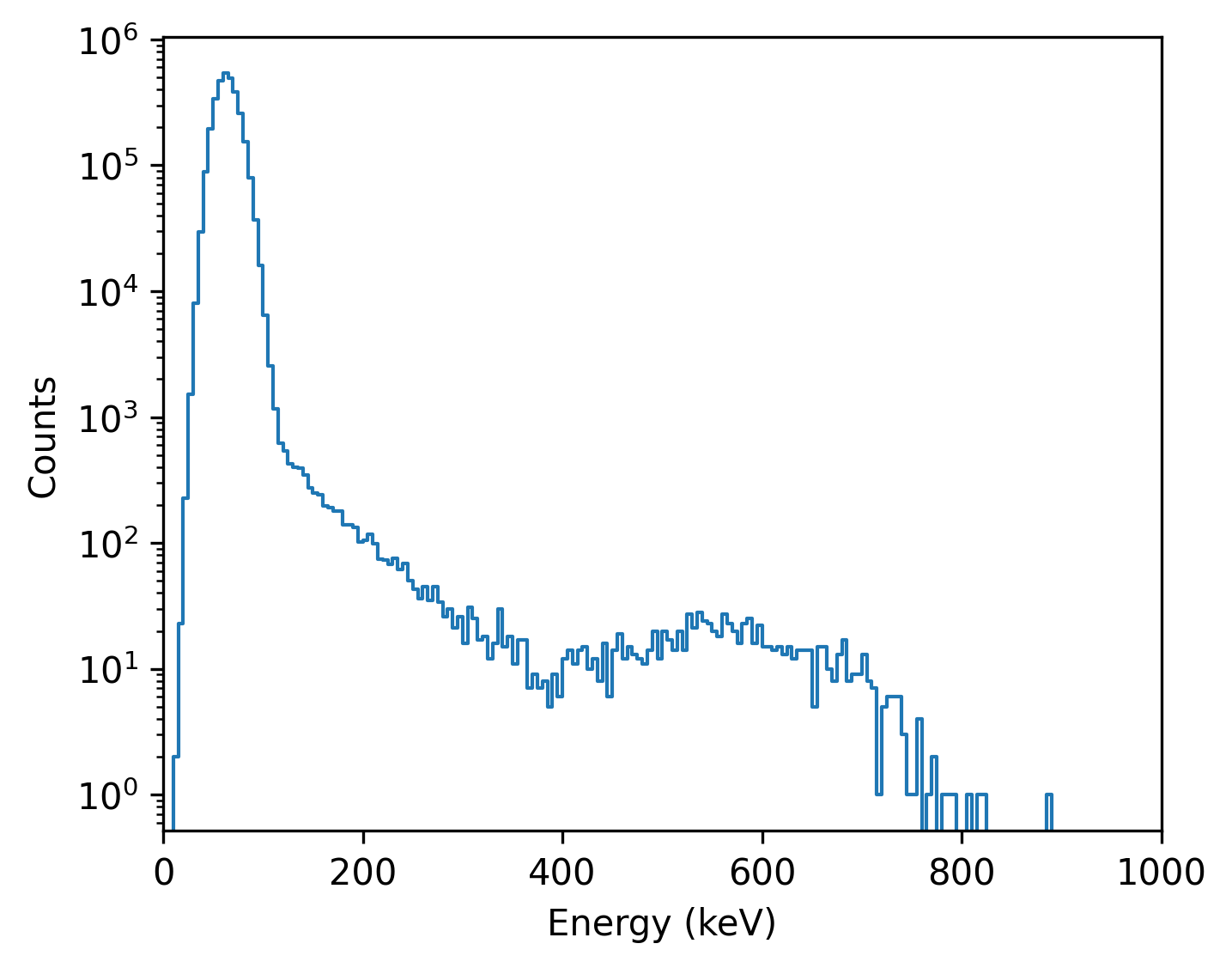}
    \caption{Example of a minute-long background spectral measurement from 23 April 2025 taken by GRBAlpha close to the equator and far from the SAA. The peak around 50~keV is due to thermal noise produced by the SiPMs.}
    \label{fig:ex_spectrum}
\end{figure}

The detected particles/$\gamma$-rays, recorded as counts, are binned on board both in time and energy.
The pulse heights, which are related to the measured energy, can be binned into $2^M$, $M=\{1,..,8\}$, equally wide bands in 
analog-to-digital units 
(ADU). The finest spectral resolution of 256 
analog-to-digital (A/D) converter channels
was used for calibration measurements (Fig.~\ref{fig:ex_spectrum}). During nominal observations we suppressed the energy channels with significant thermal noise by defining a~low-energy cutoff $ch_{\rm cut}$. This value was found empirically from the calibration measurements.
All GRBAlpha measurements presented in this work had $ch_{\rm cut}$=54~ADU. For VZLUSAT-2, the initial measurements had $ch_{\rm cut}$=45~ADU and in September 2022 it changed to $ch_{\rm cut}$=48~ADU.

The initial conversion between the pulse height amplitude  
and energy in physical units measured before the launch changed in orbit due to the SiPM degradation \citep{Ripa2019}.
For GRBAlpha, \cite{Ripa2025} revised this conversion using in-orbit measurements of activation lines caused by the high-energy protons in the South Atlantic Anomaly \citep[SAA,][]{Galgoczi2021a, Galgoczi2021b}. The revised energy conversion relation follows
\begin{equation}
    E~(keV) = g(t) \times x + E_0,
\end{equation}
where $E_0=-154\rm~keV$ is the offset value, $x$ is the spectral channel number in ADU and $g(t)$ is the gain factor which, according to \cite{Ripa2025}, changes in time as
\begin{equation} \label{eq:g}
    g(t) = a\times t^3 + b\times t^2 + c\times t + d,
\end{equation}
where the best fit parameters are $a=5.54\times 10^{-10}$, $b=-1.91\times10^{-6}$, $c=2.29\times10^{-3}$, $d=4.02$ and $t$ is the mission elapsed time in days. 
The time dependence of the gain factor for VZLUSAT-2 could not be studied due to the lack of spectral measurements after SAA passes. Therefore, we treat the gain factor as a constant and use its initial value found from ground laboratory calibrations during the entire mission. For the detector unit \#0, the conversion relation follows
\begin{equation}
    E~(keV) = 9.12 \times x - 366,
\end{equation}
while for the unit \#1 the relation was measured as
\begin{equation}
    E~(keV) = 7.84 \times x - 313.
\end{equation}

\begin{deluxetable}{rcccc}
\tablecaption{Energy bands used during regular observations and their conversion from the channels of the A/D converter (ADU) to keV.
}
\tablehead{
\colhead{} & \colhead{band 0} & \colhead{band 1} & \colhead{band 2} & \colhead{band 3}
}
\startdata
ch. (ADU) & $ch_{\rm cut}$--63 & 64--127 & 128--191 & 192--255 \\
$E_{\rm beg}$ (keV) & 60--100 & 100--360 & 360--620 & 620--870 \\
$E_{\rm end}$ (keV) & 120--170 & 170--490 & 490--810 & 810--1130 \\
$E_{\rm V0}$ (keV) & 40/70--220 & 220--800 & 800--1380 & 1380--1960 \\
$E_{\rm V1}$ (keV) & 40/60--190 & 190--700 & 700--1200 & 1200--1710 \\
\enddata
\tablecomments{$ch_{\rm cut}$ is the A/D converter channel cutoff applied to avoid the thermal noise peak. Values $E_{\rm beg}$ and $E_{\rm end}$ correspond to the beginning and end of the GRBAlpha mission. $E_{\rm V0}$ and $E_{\rm V1}$ are values for the detector unit 0 and 1 on board VZLUSAT-2, respectively.
The cutoff channel for VZLUSAT-2 has two energy values assigned. The first one is for measurements prior to September 2022 and the second one applies to data since September 2022.
}
\label{tab:energy_bands}
\end{deluxetable}

\begin{figure}
    \centering
    \includegraphics[width=\linewidth]{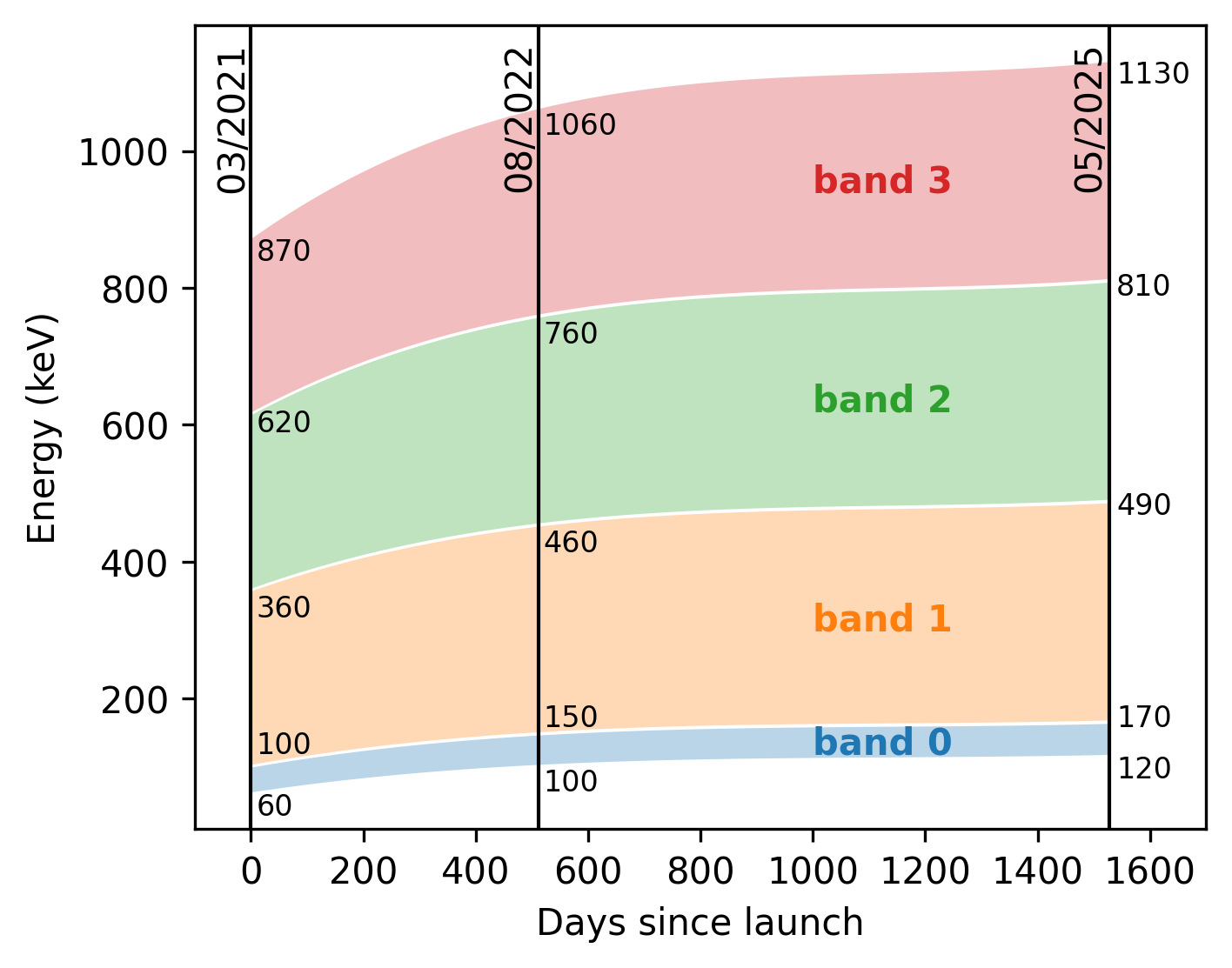}
    \caption{Change of energy bands caused by the degradation of SiPMs on board GRBAlpha. The vertical lines show specific values at three important dates; the launch of the mission in March 2021, the beginning of regular operations in August 2022, and the end of operations in May 2025.
    }
    \label{fig:ebands-change-in-time}
\end{figure}

The finest possible temporal resolution defined by the on-board firmware was 20~ms \citep{Pal2023}. However, small power budget and memory restricted us from using it for nominal measurements. 
With GRBAlpha we tested different combinations of temporal and spectral resolutions. 
Two major constraints were an available memory of $\sim2~\rm~MB$ and a variable downlink rate, depending on the number of available ground stations, typically below 500~kB/day. 
As we wanted to keep some spectral resolution to eliminate spurious events more easily, we chose to do nominal observations in four energy bands. With a resolution time of $\Delta t=1\rm~s$ we were able to store on board nearly three days of measurements which gave us enough reserve to download all relevant data.
Once we acquired a diverse sample of detected events, we decided to improve the resolution time to $\Delta t=0.5\rm~s$ to be more sensitive to short events. This limited our storage to approximately one and a half day which at times meant that we were not able to download all desired data. Routine measurements with $\Delta t=0.5\rm~s$ were done since February 2024.  
Tab.~\ref{tab:energy_bands} and Fig.~\ref{fig:ebands-change-in-time} show how the four energy bands changed in time. 
Due to the rapid degradation shortly after launch, in our comparative analysis we only include events detected after August 2022. This results in exclusion of only five GRBs detected within a period of three months in 2021. Since October 2022, GRBAlpha did nearly nonstop observations except for a~few periods when the used ground stations encountered issues.

Nominal measurements by the gamma-ray detectors on board VZLUSAT-2 were done with 1~s resolution time in four energy bands (Tab.~\ref{tab:energy_bands}) during the entire mission lifetime. As the detectors were among the satellite's secondary payloads, the frequency of observations highly depended on the data volume produced by other instruments and their power consumption. 
If an event was observed by both detector units, in the analysis presented in Sec.~\ref{sec:catalogue} and \ref{sec:fermi-gbm} we include the detector with higher significance during the entire event duration.

\subsection{Search for Transients}
\label{sec:search_for_transients}

Since GRBAlpha and VZLUSAT-2 were technological missions with a primary goal to study the performance of a new detector design, they did not provide localization nor had implemented real-time triggering.
Moreover, due to limited downlink capacity, we were not able to transmit all acquired measurements. Therefore, we followed 
trigger alerts
from other missions and downloaded only a short time interval centered around the trigger time reported on their mission’s website or via a GCN Circular\footnote{\href{https://gcn.nasa.gov/circulars/}{https://gcn.nasa.gov/circulars/}}.
This was typically one hour for GRBAlpha and five minutes for VZLUSAT-2. To identify possible observed transients, we manually inspected the measured light curves and compared them to those measured by other missions. We split the observed events into \textit{sub-threshold} detections if the signal-to-noise ratio $SNR$ falls between $3\sigma\leq SNR<5\sigma$ and \textit{significant} detections if $SNR\geq5\sigma$ (Sec.~\ref{sec:obs_properties}, Appendix~\ref{appendix:calculations}).
Therefore, the criteria for calling a spike in our light curves a ``detection'' were the $SNR$ value and the temporal coincidence with a burst reported by another mission.
At polar orbit the satellites regularly pass through the Van Allen radiation belts. Thus, for events detected near these regions we closely examined the shape of our light curves contaminated with variable background and compared them with observations from different missions. For events with known localization we further confirmed that they were not occulted by the Earth.

We followed 
trigger alerts
from past and current hard X-ray or gamma-ray missions \textit{Fermi}/GBM \citep{FermiGBM}, \textit{Swift}/BAT \citep{SwiftBAT}, \textit{INTEGRAL}/SPI \citep{IntegralSPI}, \textit{INTEGRAL}/IBIS \citep{IntegralIBIS}, \textit{Wind}/Konus \citep{WindKONUS}, \textit{AGILE} \citep{AGILE}, \textit{AstroSat} \citep{ASTROSAT}, \textit{GECAM} \citep{GECAM}, \textit{CALET}/CGBM \citep{CALET}, \textit{MAXI} \citep{MAXI}, \textit{SVOM}/GRM \citep{SVOM} and \textit{EP}/WXT \citep{EP}, as well as gravitational wave triggers from the LVK collaboration \citep{LVK} and requests for the GUANO pipeline \citep{GUANO} which include also triggers from neutrino events or fast radio bursts.

\subsection{Pile-up Correction}

\cite{Ripa2023b} estimated that for our detector the pulse pile-up becomes significant for detected count rate of $\gtrapprox2500$\,cnt\,s$^{-1}$ when the pile-up probability reaches $\sim3.7~\%$. 
Therefore, we corrected the light curves for pile-up following the method of \cite{Ripa2023b} for all events which exceed this threshold, namely GRB~221009A, GRB~230307A, GRB~250407A, and nine bright solar flares. 
Out of these, the pile-up was most significant for the brightest-of-all-time GRB 221009A when the pile-up probability reached 33~\%.

\subsection{Observational Properties} \label{sec:obs_properties}

We characterize every detection by a set of calculated parameters. These include duration, peak flux, fluence, hardness ratio, and detection significance.  
A detailed description of how the parameters and their uncertainties were calculated is provided in Appendix~\ref{appendix:calculations}.

We estimate the duration of all types of transients in terms of the $T_{50}$ and $T_{90}$ durations proposed for GRBs by \cite{Kouveliotou93}. Since most transients are only observed in the two lowest energy bands, the durations are calculated from cumulative counts measured in bands 0 and 1.
This corresponds to approximately 100--500~keV for GRBAlpha (Fig.~\ref{fig:ebands-change-in-time}) and 60--800~keV for VZLUSAT-2 (Tab.~\ref{tab:energy_bands}).

Due to the lack of attitude information of the satellites and thus impossibility to perform spectral modeling, we express the peak flux $P$ and fluence $S$ in the units of counts per second and number of counts, respectively. These
are calculated individually for band 0 (subscript~0), band 1 (subscript 1), and the sum of all four bands (no subscript). The peak time $t_p$ is determined from the entire energy range. 

The hardness ratio $HR$ is calculated as the ratio of fluence $S_1$, measured in band~1, to  $S_0$, measured in band~0 (Tab.~\ref{tab:energy_bands}). 
Note that because the fluence is defined in terms of detected counts, the measured $HR$ is direction dependent. If the incident photons enter the detector after passing through the satellite's body, more low-energy photons are absorbed along the way leading to a higher value of $HR$. 
We estimated the extent of this effect via simulations. We took
the typical spectra of short and long GRBs as measured by Fermi/GBM
\citep{GBM_spec_cat}. The spectral parameters for the cutoff power-law
model (COMP) were following: the spectral index $\alpha=-1.01$ and peak
energy $E_{\rm p}=205$\,keV for a typical long GRB and $\alpha=-0.39$
and $E_{\rm p}=532$\,keV for a typical short GRB (see also
Tab.~\ref{tab:faint_grbs}). Then we took GRBAlpha detector response
matrices (DRMs) simulated for 768 directions isotropically covering a
sphere around the satellite using Hierarchical Equal Area isoLatitude
Pixelization (HEALPix)\footnote{\url{https://healpix.sourceforge.io}}
tessellation \citep{Gorski2005}. The DRMs were simulated using GEometry
ANd Tracking (Geant4) toolkit\footnote{\url{https://geant4.web.cern.ch}}
\citep{geant4} as described in \cite{Ripa2023b}. For each direction, we
folded the GRB photon spectrum with the simulated DRM and calculated the
hardness ratio $HR=S_1/S_0$, where the fluence $S_0$ was taken as total
counts over ADC channels 54-63 and the fluence $S_1$ was taken as total
counts over channels 64-127. The simulations were done separately for
the two typical spectra of short and long GRBs. The distributions of
simulated hardness ratios were $HR=2.7_{-0.4}^{+0.3}$ (68\% CI) and
$HR=1.1\pm{0.1}$ (68\% CI) for typical short and long GRBs,
respectively. This shows that the effect of the directional dependence
of measured $HR$ calculated from the detected counts is rather minor.

The detection significance is calculated in terms of the signal-to-noise ratio $SNR_p$ for the peak flux and $SNR_s$ during the entire duration of each burst and in the full energy range (Appendix~\ref{appendix:calculations}). We define detections with $SNR\geq5\sigma$ as \textit{significant} detections and those with $3\sigma\leq SNR\leq5\sigma$ as \textit{sub-threshold} detections. This division is done separately for the peak flux and fluence.
A few transients were observed very faintly, either only with good SNR in one energy band or only at the peak. These are typically events with low fluence observed in higher background. If at least one of $SNR_p$ and $SNR_s$ satisfies the aforementioned conditions, and the detection time and shape of the light curve is consistent with other missions, we keep the event as a valid detection.

\section{The Catalogue} \label{sec:catalogue}

\begin{figure}
    \centering
    \includegraphics[width=\linewidth]{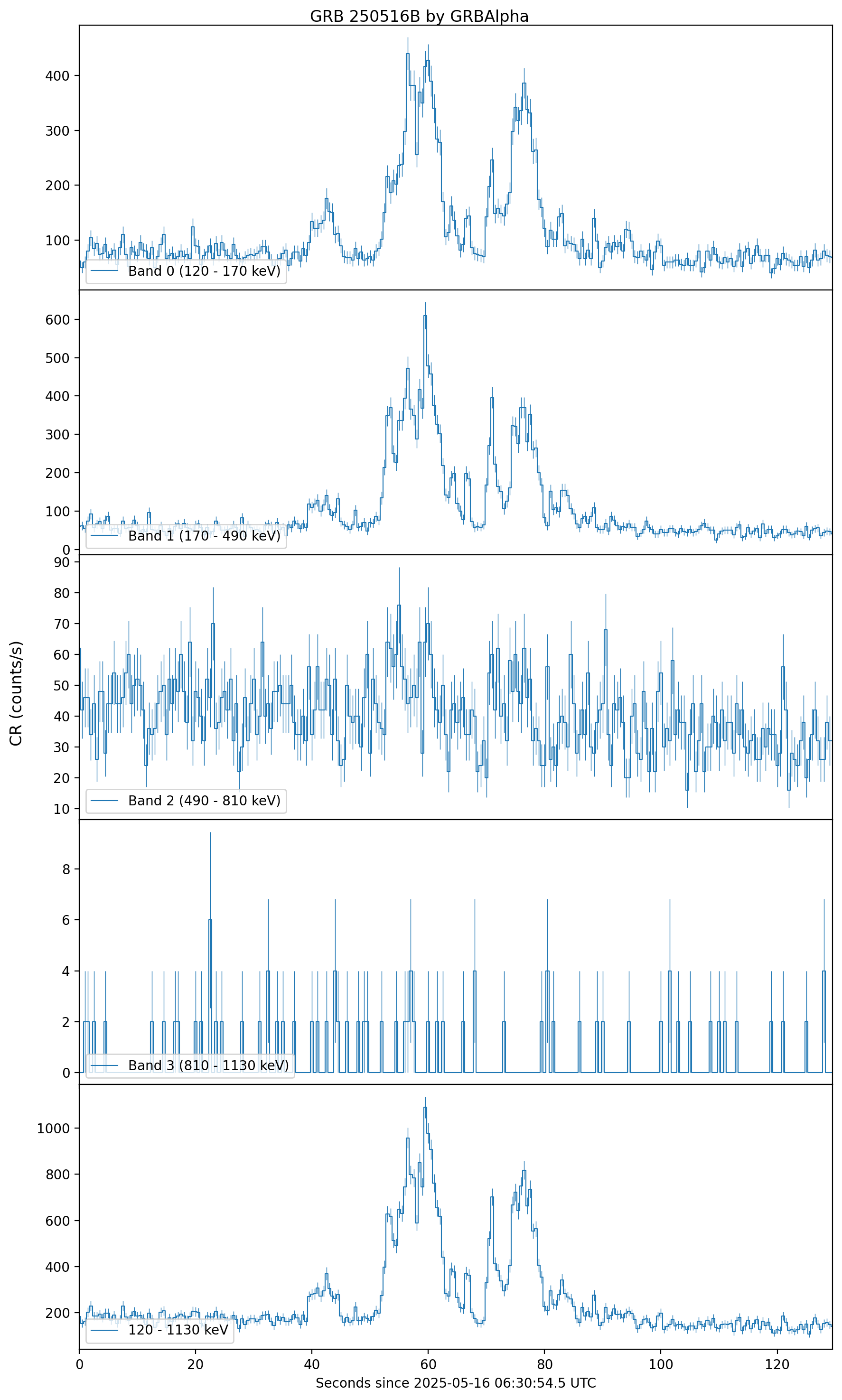}
    \caption{Example light curve of GRB 250516B observed by GRBAlpha in four energy bands and their sum. Light curves of all detected events are available in the online journal.
    }
    \label{fig:sample_lc}
\end{figure}

The catalogue contains 344 gamma-ray transients detected by the GRBAlpha and VZLUSAT-2 CubeSats. 
The majority of these are GRBs and solar flares, however, 
five bursts from magnetar SGR 1935+2154 \citep{SGR1935+2154} and one from SGR 1806-20 \citep{SGR1806-20} were observed as well as one outburst from an X-ray binary system LS V +44 17 / RX J0440.9+4431 \citep{LS_RX_1,LS_RX_2}.
Out of the 173 GRBs, 17 were only detected by missions without localization capability and were not localized by the InterPlanetary Network (IPN)\footnote{\href{https://heasarc.gsfc.nasa.gov/docs/heasarc/missions/ipn.html}{https://heasarc.gsfc.nasa.gov/docs/heasarc/missions/ipn.html}}. Independent observations by GRBAlpha and VZLUSAT-2 rule out that these events were triggered due to local particle environment and place constraints on their localizations. Tab.~\ref{tab:detections_summary} summarizes the number of events detected by each mission. 
Twenty nine events were observed jointly by both satellites. 
An example of a GRB light curve in different energies is presented in Fig~\ref{fig:sample_lc}. Light curves of all events are available as a figure set in the online 
version of the article.
Preliminary analysis including raw data and references to detections by other missions was regularly reported at corresponding websites of transients observed by GRBAlpha\footnote{\href{https://monoceros.physics.muni.cz/hea/GRBAlpha/}{https://monoceros.physics.muni.cz/hea/GRBAlpha/}} and VZLUSAT-2\footnote{\href{https://monoceros.physics.muni.cz/hea/VZLUSAT-2/}{https://monoceros.physics.muni.cz/hea/VZLUSAT-2/}} as well as via the GCN Circulars. 
Final calculated properties are available in machine readable tables whose structure and content are shown in Tab.~\ref{tab:grbalpha_dets} and \ref{tab:vzlusat2_dets}. 
Median values of the main properties for each transient class are summarized in Tab.~\ref{tab:catalogue_statistics}.

\begin{deluxetable}{cccccc}
    \tablecaption{Summary of detected transients.}    
    \tablehead{
    \colhead{mission} & \colhead{sGRB} & \colhead{lGRB} & \colhead{SF} & \colhead{SGR} & \colhead{RX} 
    }
    \startdata
    GRBAlpha & 20 & 104 & 100 & 2 & 1 \\
    VZLUSAT-2 & 8 & 60 & 73 & 5 & 0 \\
    joint detections & 1 & 18 & 9 & 1 & 0 \\
    \enddata
    \tablecomments{
    sGRB: short GRBs; lGRB: long GRBs; SF: solar flares; SGR: soft gamma repeaters; RX: X-ray binary outbursts 
    }
    \label{tab:detections_summary}
\end{deluxetable}

\begin{deluxetable}{ccccccccccccccc}
\label{tab:grbalpha_dets}
\rotate

\tabletypesize{\scriptsize}

\tablecaption{Observed characteristics of transients detected by GRBAlpha.}

\tablehead{\colhead{name} & \colhead{cutoff} & \colhead{$\Delta t$} & \colhead{$t_{\rm p}$} & \colhead{$T_{90}$} & \colhead{$T_{50}$} & \colhead{$P_0$} & \colhead{$P_1$} & \colhead{$P$} & \colhead{$S_0$} & \colhead{$S_1$} & \colhead{$S$} & \colhead{$HR$} & \colhead{$SNR_{\rm p}$} & \colhead{$SNR_{\rm s}$} \\ 
\colhead{} & \colhead{(ADU)} & \colhead{(s)} & \colhead{} & \colhead{(s)} & \colhead{(s)} & \colhead{(cnt/s)} & \colhead{(cnt/s)} & \colhead{(cnt/s)} & \colhead{(cnt)} & \colhead{(cnt)} & \colhead{(cnt)} & \colhead{} & \colhead{} & \colhead{} } 

\startdata
GRB 210807A & 54 & 4 & 2021-08-07 10:06:00 & 64$^{+9}_{-20}$ & 36$^{+6}_{-6}$ & 17$\pm$5 & 31$\pm$5 & 68$\pm$8 & 215$\pm$60 & 910$\pm$80 & 3000$\pm$200 & 4$\pm$1 & 9.0 & 16 \\
GRB 210822A & 54 & 4 & 2021-08-22 09:18:26 & 12$^{+4}_{-4}$ & 12$^{+4}_{-4}$ & 172$\pm$9 & 265$\pm$10 & 441$\pm$15 & 1300$\pm$100 & 2300$\pm$100 & 3700$\pm$200 & 1.8$\pm$0.2 & 30 & 17 \\
GRB 210909A & 54 & 4 & 2021-09-09 10:43:19 & 4$^{+4}_{-4}$ & 4$^{+4}_{-4}$ & 13$\pm$5 & 30$\pm$5 & 41$\pm$7 & 50$\pm$40 & 120$\pm$30 & 160$\pm$50 & 2.3$\pm$1.8 & 5.5 & 3.6 \\
GRB 211018A & 54 & 4 & 2021-10-18 22:29:34 & 116$^{+12}_{-6}$ & 68$^{+4}_{-4}$ & 82$\pm$6 & 167$\pm$7 & 256$\pm$10 & 5700$\pm$200 & 9500$\pm$160 & 15800$\pm$300 & 1.66$\pm$0.06 & 26 & 61 \\
GRB 211019A & 54 & 4 & 2021-10-19 05:59:34 & 40$^{+6}_{-9}$ & 20$^{+4}_{-6}$ & 35$\pm$5 & 158$\pm$8 & 229$\pm$10 & 830$\pm$100 & 2200$\pm$100 & 3600$\pm$200 & 2.7$\pm$0.4 & 22 & 21 \\
SF 220815 & 54 & 4 & 2022-08-15 14:35:06 & 100$^{+9}_{-9}$ & 60$^{+4}_{-4}$ & 141$\pm$7 & 19$\pm$5 & 158$\pm$9 & 4400$\pm$140 & 360$\pm$130 & 4600$\pm$200 & 0.08$\pm$0.03 & 17 & 22 \\
SF 220816 & 54 & 4 & 2022-08-16 00:08:38 & 48$^{+9}_{-6}$ & 24$^{+6}_{-4}$ & 40$\pm$5 & -5$\pm$3 & 40$\pm$7 & 1130$\pm$90 & 80$\pm$70 & 1400$\pm$130 & 0.07$\pm$0.06 & 6.2 & 11 \\
GRB 220826B & 54 & 3 & 2022-08-26 10:21:20 & 12$^{+3}_{-4}$ & 9$^{+3}_{-4}$ & 21$\pm$5 & 13$\pm$5 & 39$\pm$9 & 170$\pm$40 & 90$\pm$40 & 340$\pm$60 & 0.53$\pm$0.25 & 4.3 & 5.4 \\
GRB 220829A & 54 & 3 & 2022-08-29 14:37:50 & 12$^{+4}_{-3}$ & 6$^{+4}_{-3}$ & 55$\pm$6 & 56$\pm$7 & 123$\pm$11 & 600$\pm$100 & 325$\pm$90 & 900$\pm$150 & 0.55$\pm$0.18 & 12 & 5.8 \\
SF 220904 & 54 & 3 & 2022-09-04 12:13:29 & 15$^{+4}_{-4}$ & 6$^{+4}_{-3}$ & 36$\pm$5 & 10$\pm$4 & 46$\pm$7 & 280$\pm$60 & 90$\pm$60 & 350$\pm$80 & 0.3$\pm$0.2 & 6.3 & 4.6 \\
\enddata

\tablecomments{Full version of this table is available online in machine-readable form. Transients other than GRBs are named analogously to the GRB convention.}

\end{deluxetable}

\begin{deluxetable}{cccccccccccccccc}
\label{tab:vzlusat2_dets}

\rotate

\tabletypesize{\scriptsize}

\tablecaption{Observed characteristics of transients detected by VZLUSAT-2.}

\tablehead{\colhead{name} & \colhead{det. unit} & \colhead{cutoff} & \colhead{$\Delta t$} & \colhead{$t_{p}$} & \colhead{$T_{90}$} & \colhead{$T_{50}$} & \colhead{$P_0$} & \colhead{$P_1$} & \colhead{$P$} & \colhead{$S_0$} & \colhead{$S_1$} & \colhead{$S$} & \colhead{$HR$} & \colhead{$SNR_{p}$} & \colhead{$SNR_{s}$} \\ 
\colhead{} & \colhead{} & \colhead{(ADU)} & \colhead{(s)} & \colhead{} & \colhead{(s)} & \colhead{(s)} & \colhead{(cnt/s)} & \colhead{(cnt/s)} & \colhead{(cnt/s)} & \colhead{(cnt)} & \colhead{(cnt)} & \colhead{(cnt)} & \colhead{} & \colhead{} & \colhead{} } 

\startdata
GRB 220320A & 1 & 45 & 15 & 2022-03-20 04:39:55 & 30$^{+15}_{-15}$ & 15$^{+21}_{-15}$ & 86$\pm$4 & 92$\pm$3 & 181$\pm$6 & 2000$\pm$170 & 1700$\pm$100 & 3800$\pm$240 & 0.87$\pm$0.09 & 30 & 16 \\
SF 220421 & 0 & 45 & 1 & 2022-04-21 21:03:46 & 48$^{+17}_{-26}$ & 16$^{+5}_{-7}$ & 100$\pm$10 & 4$\pm$7 & 110$\pm$20 & 1600$\pm$140 & -320$\pm$20 & 1600$\pm$200 & -0.2$\pm$0.02 & 6.5 & 8.1 \\
SF 220421 & 1 & 45 & 1 & 2022-04-21 21:03:47 & 62$^{+5}_{-6}$ & 21$^{+2}_{-2}$ & 350$\pm$20 & 12$\pm$7 & 370$\pm$20 & 6050$\pm$150 & 240$\pm$80 & 6000$\pm$200 & 0.04$\pm$0.01 & 16 & 33 \\
GRB 220423A & 0 & 45 & 1 & 2022-04-23 14:14:14 & 31$^{+13}_{-14}$ & 9$^{+2}_{-2}$ & 110$\pm$10 & 190$\pm$15 & 320$\pm$20 & 900$\pm$130 & 1130$\pm$100 & 2100$\pm$200 & 1.2$\pm$0.2 & 14 & 12 \\
GRB 220423A & 1 & 45 & 1 & 2022-04-23 14:14:14 & 20$^{+16}_{-5}$ & 8$^{+1}_{-1}$ & 125$\pm$15 & 200$\pm$20 & 325$\pm$20 & 830$\pm$120 & 1100$\pm$90 & 2000$\pm$200 & 1.3$\pm$0.2 & 15 & 12 \\
SF 220520 & 1 & 45 & 1 & 2022-05-20 22:09:59 & 10$^{+3}_{-2}$ & 4$^{+1}_{-1}$ & 60$\pm$10 & 13$\pm$7 & 80$\pm$16 & 320$\pm$80 & 30$\pm$45 & 340$\pm$90 & 0.09$\pm$0.15 & 5.1 & 3.6 \\
GRB 220608B & 1 & 45 & 1 & 2022-06-08 07:36:39 & 17$^{+5}_{-6}$ & 9$^{+2}_{-3}$ & 40$\pm$10 & 21$\pm$8 & 70$\pm$15 & 18-$\pm$90 & 130$\pm$60 & 500$\pm$115 & 0.7$\pm$0.5 & 4.6 & 4.4 \\
SF 220715 & 1 & 45 & 1 & 2022-07-15 23:08:47 & 36$^{+5}_{-5}$ & 12$^{+2}_{-1}$ & 150$\pm$20 & -3$\pm$6 & 150$\pm$20 & 2500$\pm$200 & 10$\pm$30 & 2300$\pm$200 & 0.004$\pm$0.012 & 7.4 & 10 \\
GRB 220719A & 1 & 45 & 1 & 2022-07-19 19:09:43 & 8$^{+1}_{-1}$ & 5$^{+1}_{-2}$ & 65$\pm$15 & 31$\pm$8 & 100$\pm$20 & 190$\pm$90 & 100$\pm$40 & 310$\pm$100 & 0.5$\pm$0.3 & 5.2 & 3.1 \\
GRB 220912A & 1 & 48 & 1 & 2022-09-12 00:50:23 & 33$^{+6}_{-2}$ & 17$^{+2}_{-1}$ & 55$\pm$10 & 47$\pm$9 & 100$\pm$15 & 1000$\pm$120 & 765$\pm$85 & 1700$\pm$160 & 0.8$\pm$0.1 & 6.7 & 11 \\
\enddata

\tablecomments{Full version of this table is available online in machine-readable form. Transients other than GRBs are named analogously to the GRB convention.}

\end{deluxetable}

\subsection{Detection Rate} \label{sec:det_rate}

\begin{figure}
    \centering
    \includegraphics[width=\linewidth]{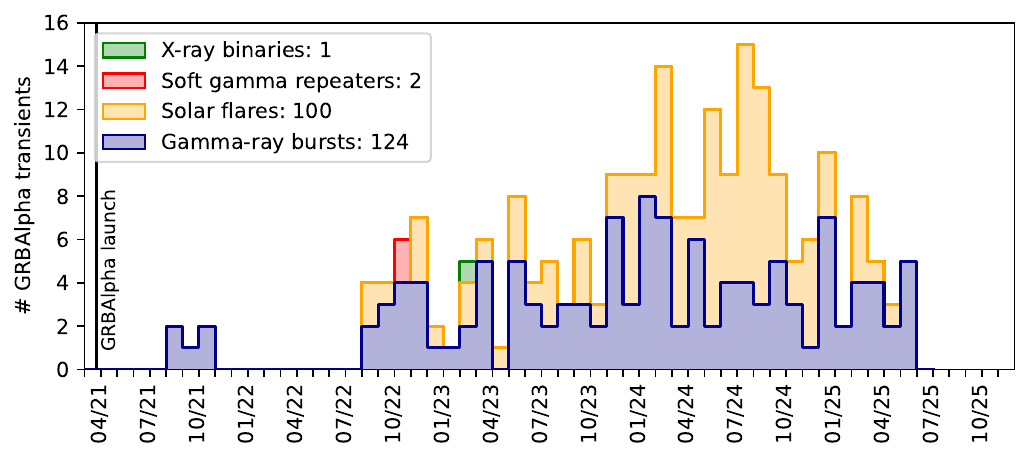}

    \vspace{0.5cm}

    \includegraphics[width=\linewidth]{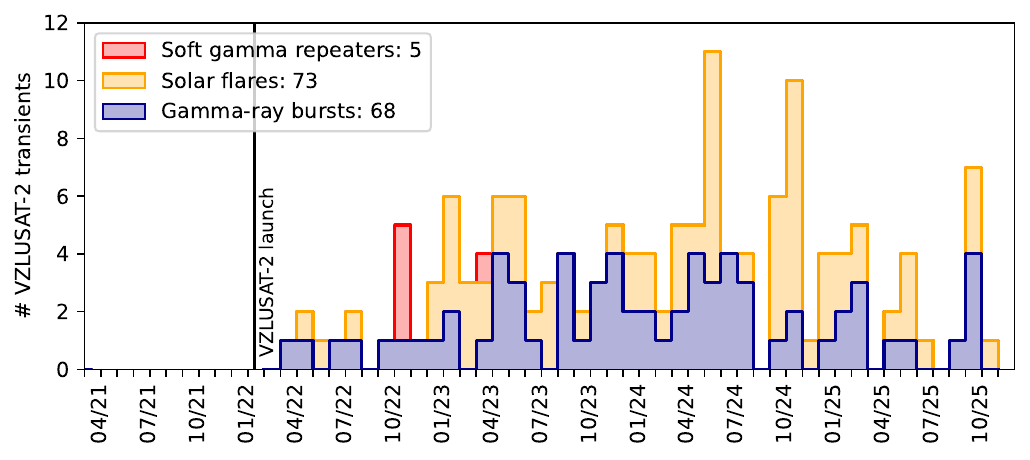}

    \caption{Stacked bar charts showing the monthly detection rate of transients observed by GRBAlpha (top) and VZLUSAT-2 (bottom).}
    \label{fig:det_rate_both}
\end{figure}

GRBAlpha detected 227 transients during its entire lifetime of more than four years. The top part of Fig.~\ref{fig:det_rate_both} shows the monthly detection rate since its launch. During the first 16 months of the mission, the satellite only detected five GRBs. This was mainly because of limited on-board memory and downlink rate but also due to the damage of one of two radio transceivers in October 2021 which caused significant issues with the communication. Thanks to a major firmware upgrade, improvements in our ground segment, use of the amateur radio network SatNOGS\footnote{\href{https://network.satnogs.org/}{https://network.satnogs.org/}} and invaluable time of the operators, GRBAlpha was able to run nearly nonstop observations since October 2022 until less than three weeks before re-entry on 24 May 2025.
During this time, there were several interruptions in the routine measurements lasting up to a few weeks due to ground station issues.
Considering only periods when measurements were ongoing continuously for at least one week, GRBAlpha detected on average two transients, or one GRB, per week. 
The busiest month 
was July 2024 with four GRB and 11 solar flare detections, i.e., one transient was observed every two days. The shortest time between two subsequent detections was 42~minutes between GRB~230709B and GRB~230709C, followed by 47 minutes between GRB 250204A and GRB 250204B. 
Moreover, the very last transient, GRB 250521A, was observed less than three weeks before GRBAlpha re-entered the Earth's atmosphere.

VZLUSAT-2 detected 146 transients during almost four years of operation from March 2022 to November 2025. The monthly detection rate is depicted in the bottom part of Fig.~\ref{fig:det_rate_both} and corresponds to one detection every 10~days. GRB was on average observed once every three weeks, however, the shortest time between two GRBs was 106 minutes in case of GRB 240710A and GRB 240710B. 
Despite having two gamma-ray detectors on board, VZLUSAT-2 observed less transients than GRBAlpha. As VZLUSAT-2 carried more payloads, it was not possible to run measurements with our detectors continuously and the overall amount of data from VZLUSAT-2 is smaller than that from GRBAlpha.

The detection rate is not uniform in time. Bursts from SGRs and X-ray binaries were observed sporadically as their spectra are softer and only a handful of them reach high enough photon flux in energies our detector is sensitive to. 
The frequency of observed solar flares follows solar activity. Moreover, in case of GRBAlpha we sometimes sacrificed non-GRB triggers due to downlink constraints. If there were not enough ground stations available, we did not download data from times of, especially, many solar flares. 

\subsection{Observed properties}

\renewcommand{\arraystretch}{1.15} 
\begin{deluxetable}{cccccc}
\tablecaption{Median values and one standard deviation intervals of observed properties for GRBAlpha and VZLUSAT-2 detections.}
\label{tab:catalogue_statistics}
\tablehead{
\colhead{} & \colhead{$T_{50}$} & \colhead{$T_{90}$} & \colhead{$P$} & \colhead{$S$} & \colhead{$HR$} \\
\colhead{} & \colhead{(s)} & \colhead{(s)} & \colhead{(cnt/s)} & \colhead{(cnt)} & \colhead{}
}
\startdata
\multicolumn{6}{c}{{GRBAlpha}} \\
\hline
sGRB & $1.0^{+1.0}_{-0.5}$ & $1.0^{+2.0}_{-0.5}$ & $130^{+80}_{-70}$ & $130^{+200}_{-45}$ & $1.7^{+1.7}_{-0.9}$ \\
lGRB & $7^{+16}_{-4}$ & $14^{+33}_{-8}$ & $130^{+300}_{-60}$ & $900^{+2900}_{-600}$ & $1.3^{+1.0}_{-0.6}$ \\
SF   & $24_{-18}^{+80}$ & $49_{-35}^{+180}$ & $180_{-95}^{+430}$ & $2900_{-2100}^{+20500}$ & $0.15_{-0.08}^{+0.22}$ \\
SGR  & $8_{-3}^{+3}$ & $12_{-5}^{+5}$ & $160_{-70}^{+70}$ & $1500_{-800}^{+800}$ & $0.22_{-0.01}^{+0.01}$ \\
RX   & 6 & 14 & 43 & 245 & 0.14 \\
\hline
\multicolumn{6}{c}{{VZLUSAT-2}} \\
\hline
sGRB & $1.5_{-0.5}^{+0.5}$ & $3.0_{-2.0}^{+0.0}$ & $110_{-40}^{+90}$ & $165_{-60}^{+50}$ & $1.0_{-0.5}^{+0.2}$ \\
lGRB & $9_{-5}^{+11}$ & $22_{-15}^{+20}$ & $120_{-50}^{+300}$ & $1200_{-800}^{+1200}$ & $0.8_{-0.3}^{+0.5}$ \\
SF   & $16_{-9}^{+28}$ & $39_{-23}^{+60}$ & $150_{-85}^{+400}$ & $2700_{-2100}^{+14400}$ & $0.09_{-0.09}^{+0.12}$ \\
SGR  & $1.0_{-0.0}^{+1.0}$ & $1.0_{-0.0}^{+3.8}$ & $560_{-380}^{+610}$ & $530_{-300}^{+1300}$ & $0.07_{-0.07}^{+0.07}$ \\
\enddata
\end{deluxetable}

\begin{figure}
    \hfill
    \begin{minipage}[b]{\linewidth}
        \centering
        \includegraphics[width=\linewidth]{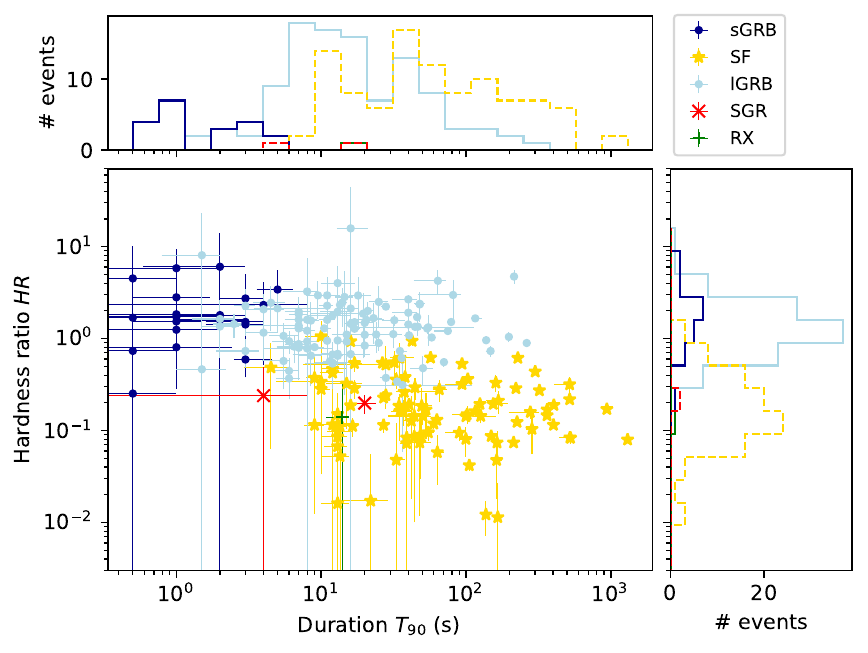}
    \end{minipage}
    \hfill
    \begin{minipage}[b]{\linewidth}
        \centering
        \includegraphics[width=\textwidth]{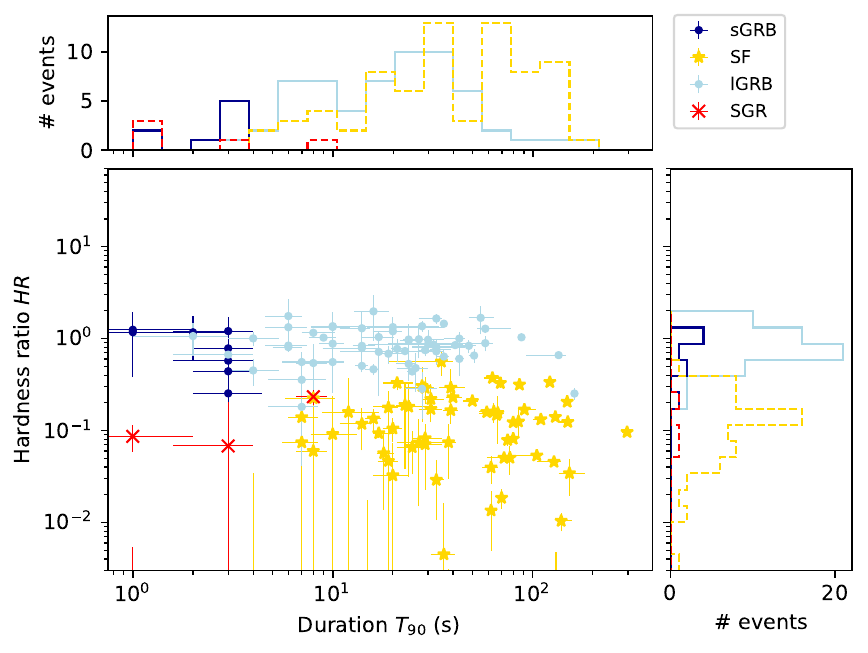}
    \end{minipage}
    \caption{Hardness ratio vs. $T_{90}$ duration of transients observed by GRBAlpha (top) and VZLUSAT-2 (bottom).
    }
    \label{fig:t90_hr}
\end{figure}

Fig.~\ref{fig:t90_hr} shows all detected transients in the $HR-T_{90}$ plane. 
GRBs are significantly harder than other classes. The bimodal distribution of their durations is not as strong as seen by other missions due to a small sample of short GRBs and limited temporal resolution. 
The division into short and long GRBs was done based on measurements done by other missions with finer temporal resolution.
The $HR-T_{90}$ plane can be roughly divided into harder GRBs spanning a~wide range of durations, softer and longer solar flares, and softer and shorter soft gamma repeaters (SGRs).
We note that nearly half of the detected solar flares, 37 from the GRBAlpha sample and 38 from VZLUSAT-2, were only detected in the lowest energy band. This means that their $HR$ is essentially the ratio of 
fluence $S1$, measured in band~1, which in those cases contains only background without clear transient detection,
to $S_0$ measured in band~0. Therefore, the absolute $HR$ value of solar flares detected only in the lowest energies may not accurately represent their characteristics. Nevertheless, the relative comparison of the full sample of solar flares to other transient classes is still correct.
Some SGR bursts appear intermediate because of a temporal resolution of 4~s, or even longer due to the so-called forest of bursts (as also reported by \cite{SGR1935+2154_forest_of_bursts}), a~collection of many brief spikes which are either indistinguishable by the temporal resolution or separated by only a few bins \citep{Israel2008,Kaneko2021}.
The one outburst from an X-ray binary detected by GRBAlpha lies among shorter solar flares, close to the boundary of different transient types.

\begin{figure*}
    \hfill
    \begin{minipage}[b]{0.49\linewidth}
        \centering
        \includegraphics[width=\linewidth]{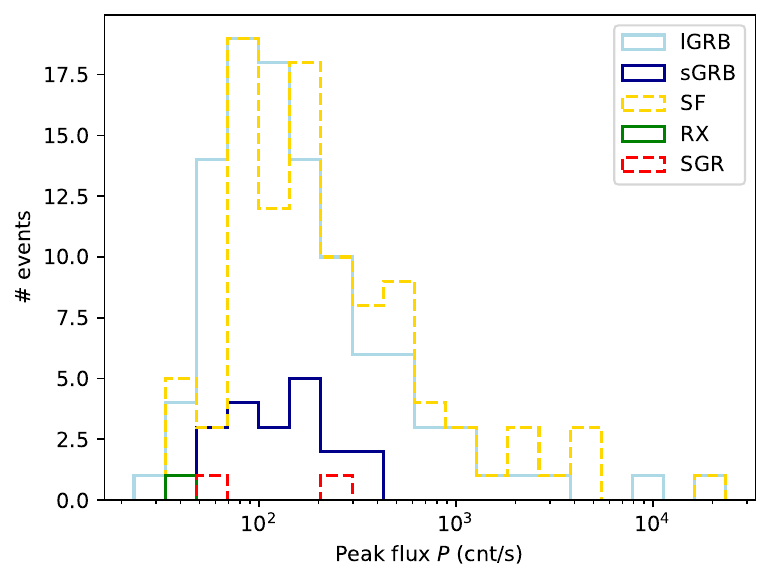}
    \end{minipage}
    \hfill
    \begin{minipage}[b]{0.49\linewidth}
        \centering
        \includegraphics[width=\textwidth]{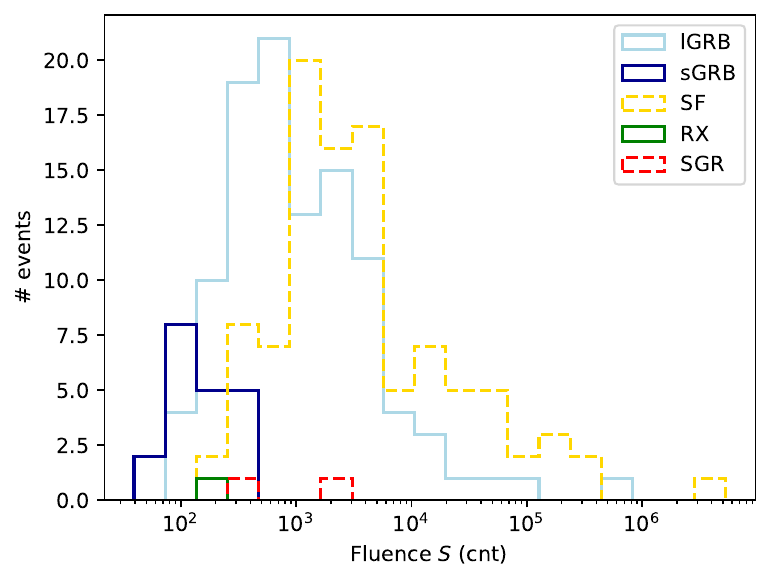}
    \end{minipage}
    \caption{Distributions of the peak flux $P$ (left) and fluence $S$ (right) for different transient types as observed by GRBAlpha. 
    }
    \label{fig:peak_flux_fluence_distributions_grbalpha}
\end{figure*}

\begin{figure*}
    \hfill
    \begin{minipage}[b]{0.49\linewidth}
        \centering
        \includegraphics[width=\linewidth]{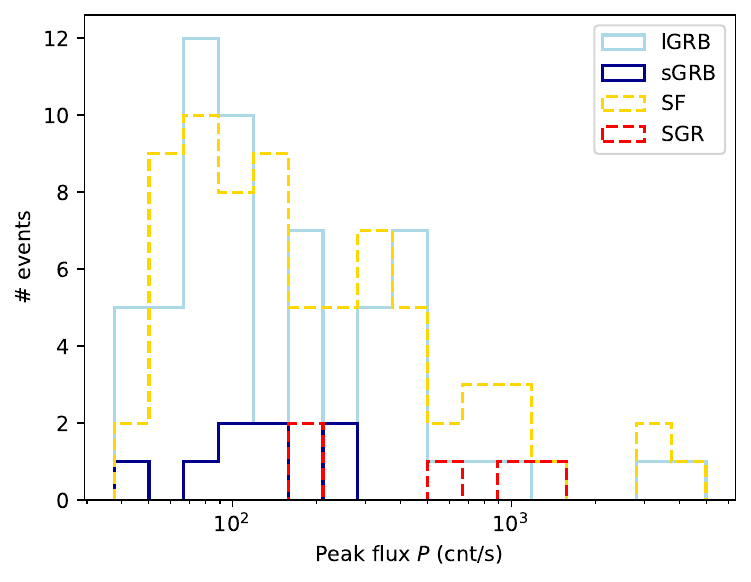}
    \end{minipage}
    \hfill
    \begin{minipage}[b]{0.49\linewidth}
        \centering
        \includegraphics[width=\textwidth]{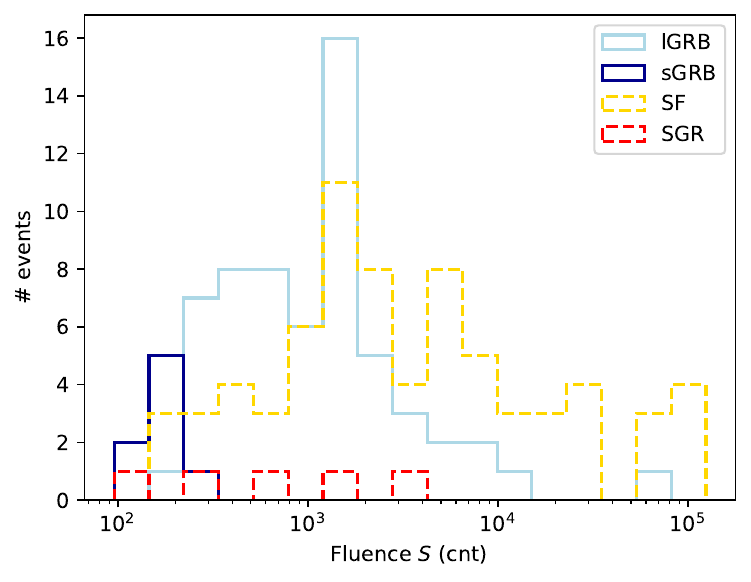}
    \end{minipage}
    \caption{Distributions of the peak flux $P$ (left) and fluence $S$ (right) for different transient types as observed by VZLUSAT-2. 
    }
    \label{fig:peak_flux_fluence_distributions_vzlusat2}
\end{figure*}

The distributions of measured peak flux and fluence are shown in Fig.~\ref{fig:peak_flux_fluence_distributions_grbalpha} and \ref{fig:peak_flux_fluence_distributions_vzlusat2}. The peak flux does not significantly vary between individual transient types. The distributions are skewed with higher median values for solar flares and SGRs. Due to the low-energy threshold we only detect the brightest transients from these classes whose luminosity peaks below our sensitivity level.
The sharper boundary at low fluxes is related to the detection limit of faint events while the weak tail toward high fluxes represents the low occurrence rate of very bright transients.
Naturally, the fluence distributions follow those of the durations. 
We note that the values of peak flux and fluence are in instrumental units of counts per second and counts.
The effective area depends on the orientation of the detector with respect to the GRB direction and is energy dependent. Simulations in \cite{Pal2023} and \cite{Ripa2023b} showed that the flat detector has minimal mean effective area when viewed from its side and it drops to about $40~\%$ for an angle of 60 degrees from the on-axis direction. As the satellite's attitude is unknown, we are not able to convert the detected counts to physical units.

When comparing the events detected jointly by GRBAlpha and VZLUSAT-2, the observed peak time is not always the same. This is primarily caused by changing attitude of the satellites. In some long-duration events which were detected by both detector units on board VZLUSAT-2, we clearly see the changing attitude in the light curves and the peak time can differ even between the two VZLUSAT-2 detector units.
Moreover, VZLUSAT-2 had issues with on-board clock synchronization between February and May 2024. Therefore, the observed peak time for transients detected during this period is shifted with respect to observations done by other missions.

\subsection{Honorable Mentions} \label{sec:events}

Among the many observed events there are a few that stand out. In this section we briefly comment on the most interesting transients detected by GRBAlpha and VZLUSAT-2.

On October 9, 2022, during a test run shortly before GRBAlpha began continuous measurements, the brightest-of-all-time \citep[BOAT,][]{Burns2023, Atteia2025} GRB 221009A blinded most of the large gamma-ray missions \citep{Frederiks2023, Lesage2023, Savchenko2024, Zhang2025}. 
Thanks to the small detector size, GRBAlpha provided an unsaturated measurement of the brightest episode of the event, although the measured light curve had to be corrected for non-negligible pulse pile-up. 
This allowed us to calculate the lower limit of its peak flux \citep{Ripa2023b}.
The only other missions which 
provided similar unsaturated measurements were GECAM-C \citep{An2023} and SRG/ART-XC \citep{Frederiks2023}. 
The S1 detector of the Konus instrument on board the Wind spacecraft also did not get saturated as it observed the burst through the satellite body, however, the complex spacecraft structure disabled the use of the data for direct calculations \citep{Frederiks2023}.

\begin{figure}
    \centering
    \includegraphics[width=\linewidth]{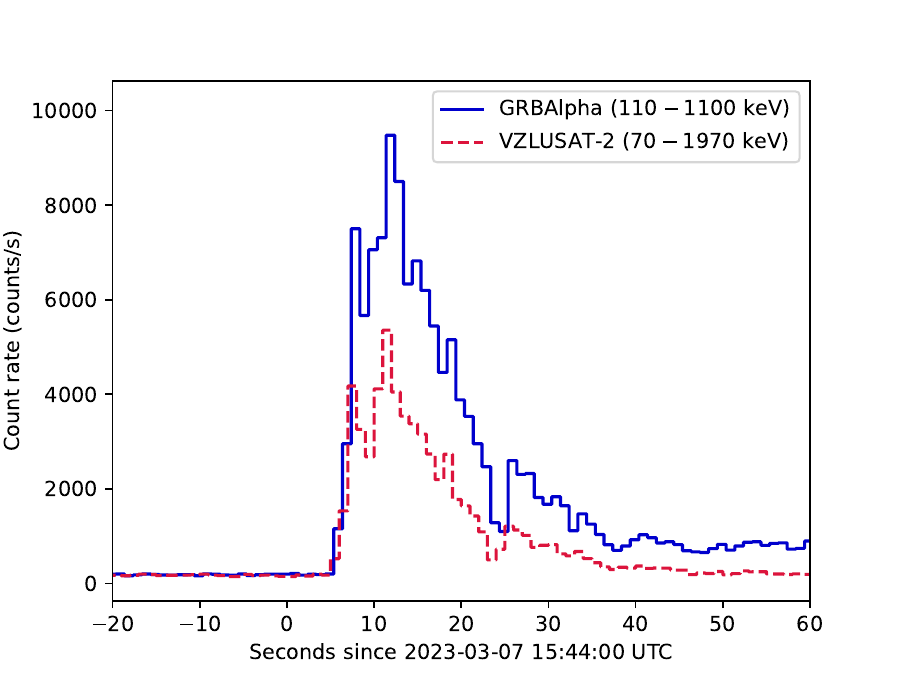}
    \caption{Light curves of GRB 230307A observed jointly by GRBAlpha and VZLUSAT-2.
    }
    \label{fig:joint_lc_grbalpha_vz2}
\end{figure}

Both GRBAlpha and VZLUSAT-2 also observed the second brightest GRB 230307A \citep{Dalessi2025, GRBAlpha_GRB230307A_GCN, Dichiara2023, Mereghetti2023, Moradi2024, Ripa2023a, Wang2023}. The measured light curves are depicted in Fig. \ref{fig:joint_lc_grbalpha_vz2}. Despite the long duration of the prompt gamma emission with $T_{90} \approx 35$\,s, this event has been associated with a kilonova detection pointing to a compact stellar merger origin \citep{Dai2024, Gillanders2025MNRAS, Levan2024Natur}. 

Although identification of GRB host galaxies is still challenging today and most GRBs lack redshift measurements, the redshift was determined for 24 GRBs included in this catalogue. Out of these, the most distant was GRB 241025A observed by VZLUSAT-2 and localized at $z=4.20$ by the Nordic Optical Telescope \citep{GRB241025A_redshift}. Assuming a flat $\Lambda$CDM model \citep[$H_0$ = 69.6~km/s/Mpc, $\Omega_{\rm M}$ = 0.29, $\Omega_0$ = 0.71,][]{cosmological_parameters}, this equals to a light travel time of 12.2~Gyr. The furthest GRB detected by GRBAlpha was GRB 241026A at $z=2.79$ \citep{GRB241026A_redshift_a,GRB241026A_redshift_b} which is equivalent to a look-back time of 11.4 Gyr. 

As a technological pathfinder toward a larger satellite constellation, GRBAlpha by itself is not capable of localizing the observed events. Nevertheless, in symbiosis with other missions within IPN, GRBAlpha was able to place further constraints on the localization of GRB 250516B \citep{Svinkin2025}.

GRBAlpha also observed the GRB 250322B which was simultaneously detected by AstroSat \citep{GRB250322B_AstroSatCZTI,GRB250322B_AstroSatLAXPC}, NuSTAR \citep{GRB250322B_NuSTAR}, and Konus-Wind\footnote{\href{https://www.ioffe.ru/LEA/kw/triggers/2025/kw20250322_72435.html}{https://www.ioffe.ru/LEA/kw/triggers/2025/kw20250322\_72435.html}}. This relatively long and bright burst shows quasi-periodic dips \citep{GRB250322B_NuSTAR} during its prompt phase which was reported by the aforementioned missions and is also seen in the light curve measured by GRBAlpha. 
Such periodicity is common for tails of SGR flares rather than GRBs. However, in this case the apparent period of about 20~s is higher than typical rotation period of magnetars.

\section{Cross-correlation with Fermi/GBM}
\label{sec:fermi-gbm}

The majority of GRBs observed by our missions were triggered by Fermi/GBM, specifically, 90 of those observed by GRBAlpha and 52 by VZLUSAT-2. In this section, we take a closer look at this subsample in order to learn about the detector sensitivity. 
The GBM data used in this section were obtained from the online Fermi GBM Burst Catalog\footnote{\href{https://heasarc.gsfc.nasa.gov/W3Browse/fermi/fermigbrst.html}{https://heasarc.gsfc.nasa.gov/W3Browse/fermi/fermigbrst.html}} \citep[FERMIGBRST,][]{Gruber2014,vonKienlin2014,NarayanaBhat2016,vonKienlin2020}, unless stated otherwise.

\subsection{Durations}

\begin{figure}
    \centering
    \includegraphics[width=\linewidth]{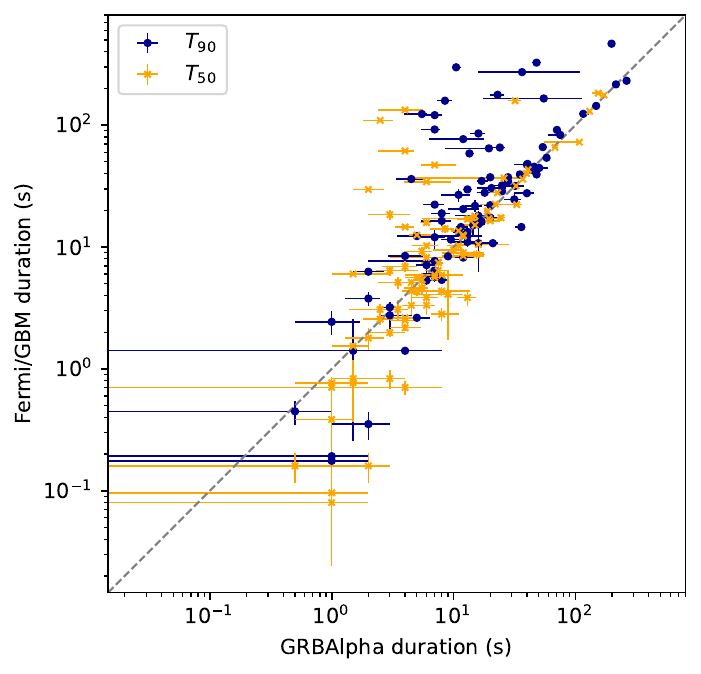}
    \caption{Comparison of the $T_{90}$ and $T_{50}$ durations of GRBs observed jointly by GRBAlpha and Fermi/GBM. 
    } 
    \label{fig:durations_comp_grbalpha_gbm}
\end{figure}

\begin{figure}
    \centering
    \includegraphics[width=\linewidth]{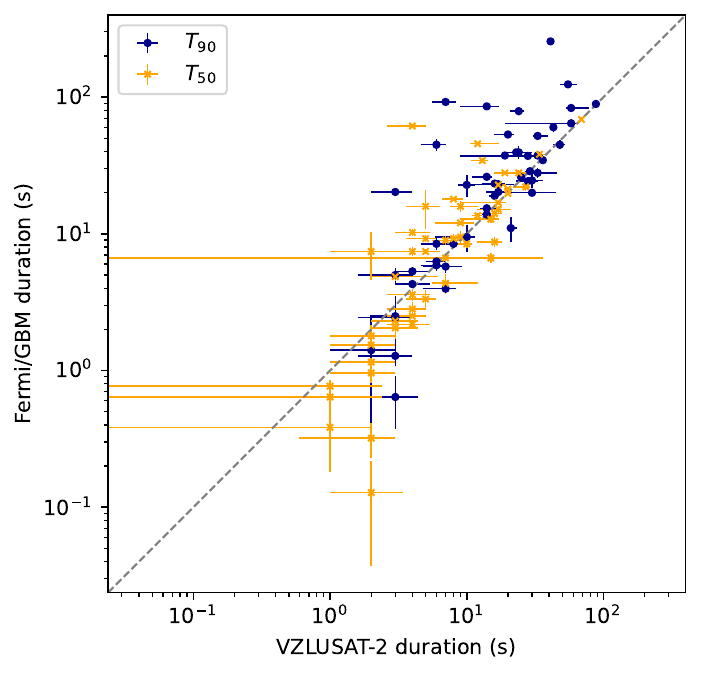}
    \caption{Comparison of the $T_{90}$ and $T_{50}$ durations of GRBs observed jointly by VZLUSAT-2 and Fermi/GBM. 
} 
    \label{fig:durations_comp_vzlusat2_gbm}
\end{figure}

Fig.~\ref{fig:durations_comp_grbalpha_gbm}, \ref{fig:durations_comp_vzlusat2_gbm} and Tab.~\ref{tab:grbalpha_vzlusat2_gbm_durations} compare the $T_{90}$ and $T_{50}$ durations measured by the CubeSat missions with Fermi/GBM. 
The GBM durations were calculated over 50--300 keV, while in our analysis we calculated durations in a combined band~0 + band~1 that goes up to 500~keV for GRBAlpha and 800~keV for VZLUSAT-2 (Tab.~\ref{tab:energy_bands}, Appendix~\ref{appendix:calculations}).
Note, that the effective area falls below 20~cm$^2$ above 300~keV \citep{Ripa2023b}.
On average, the GBM durations are longer than those measured by GRBAlpha and VZLUSAT-2. 
This agrees with the findings of \cite{Fenimore1995} that GRBs are observed longer in lower energies.
The GBM detectors also have much larger effective area and are thus more sensitive to GRB precursors and fainter extended emission. This can also explain the smaller difference in the $T_{50}$ which only encompasses the brightest part of the burst. 
Moreover, some GRBs were observed only partially by the CubeSats due to the proximity of radiation belts. For instance, the second part of the BOAT GRB~221009A was detected inside of the radiation belt \citep{GRBAlpha_GRB221009A_GCN,Ripa2023b} and the extremely bright GRB~230307A \citep{GRBAlpha_GRB230307A_GCN} was detected by GRBAlpha shortly before it entered the radiation belt and therefore the tail part of the burst is buried under the particle background. Many events were also observed near the poles where the background is usually stable but nearly twice as high as at the equator. Therefore, the significance and also the duration of these events is measured to be lower by our CubeSats than by other missions with better signal-to-noise ratio.
For short events, the sensitivity is limited by the temporal resolution.

\begin{deluxetable}{rcccc}
\label{tab:grbalpha_vzlusat2_gbm_durations}
\tablecaption{Median durations of GRBs detected simultaneously by Fermi/GBM and CubeSat mission GRBAlpha or VZLUSAT-2.
}
\tablehead{
\colhead{} & \multicolumn{2}{c}{GRBAlpha} & \multicolumn{2}{c}{VZLUSAT-2} \\
\cmidrule(lr){2-3} \cmidrule(lr){4-5}
\colhead{} & \colhead{$T_{50}$ (s)} & \colhead{$T_{90}$ (s)} & \colhead{$T_{50}$ (s)} & \colhead{$T_{90}$ (s)}
}
\startdata
        CubeSat & 6 & 14 & 7 & 17 \\
        Fermi/GBM & 8 & 22 & 9 & 24 \\
\enddata
\end{deluxetable}

\subsection{Peak Flux and Fluence}
\label{sec:gbm_flux}

\begin{figure*}
    \hfill
    \begin{minipage}[b]{0.49\linewidth}
        \centering
        \includegraphics[width=\linewidth]{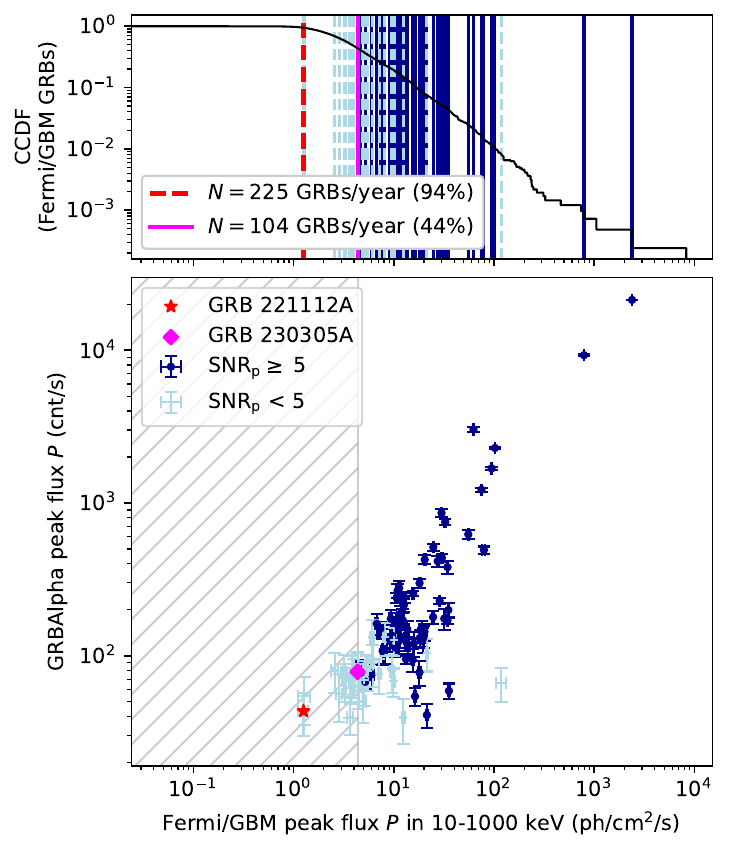}
    \end{minipage}
    \hfill
    \begin{minipage}[b]{0.49\linewidth}
        \centering
        \includegraphics[width=\textwidth]{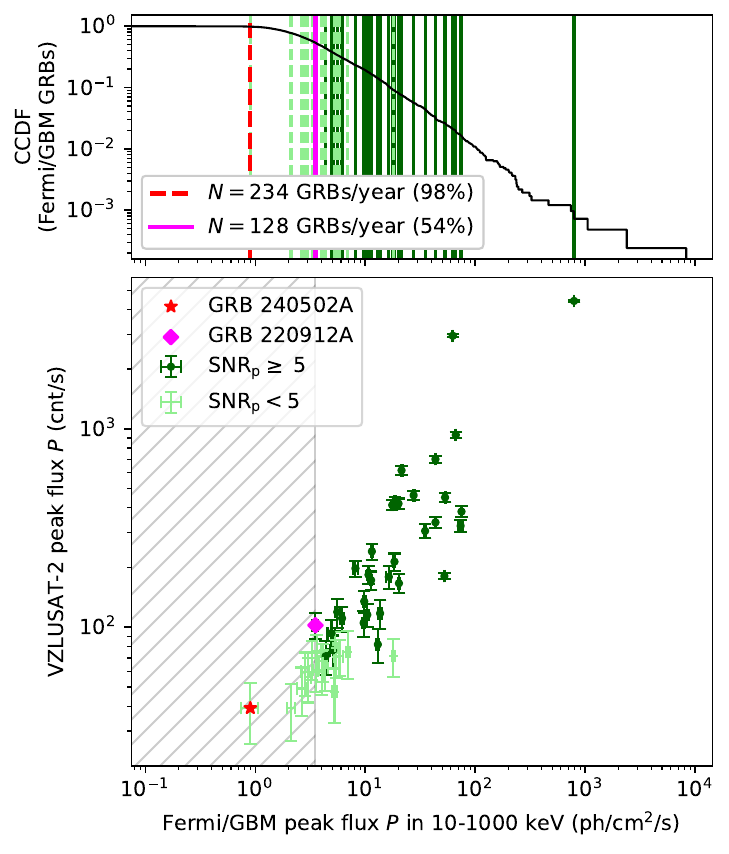}
    \end{minipage}
    \caption{
    Comparison of the peak flux $P$ of GRBs observed by GRBAlpha (left) and VZLUSAT-2 (right) with Fermi/GBM. \textit{Top:} Complementary cumulative distribution function of all GBM GRBs detected until 18 November 2025. The black curve marks the normalized number of GBM GRBs with $P$ higher than or equal to the given value.
    The solid (dashed) vertical lines represent significant (sub-threshold) detections by GRBAlpha or VZLUSAT-2. The dashed red line marks the faintest GRB from the full sample of GRBAlpha or VZLUSAT-2 GRBs. The solid magenta line is the faintest GRB out of the ones observed at a level higher than $5\sigma$. The average number of GRBs $N$ detected per year by GBM with $P$ higher than or equal to the faintest sub-threshold and significant detection by each CubeSat is stated in the legend along with the percentage of GBM GRBs this value corresponds to. 
    \textit{Bottom:} Comparison of the peak flux measured by GRBAlpha (left) and VZLUSAT-2 (right) with Fermi/GBM. The red star highlights the faintest GRB from the full sample. The magenta diamond highlights the faintest GRB from the sample of significant GRBs ($SNR\geq5\sigma$).
    The gray hatched part to the left marks a region where observed GRBs are fainter than the faintest significant GRB.}
    \label{fig:flux_gbm}
\end{figure*}

\begin{figure*}
    \hfill
    \begin{minipage}[b]{0.49\linewidth}
        \centering
        \includegraphics[width=\linewidth]{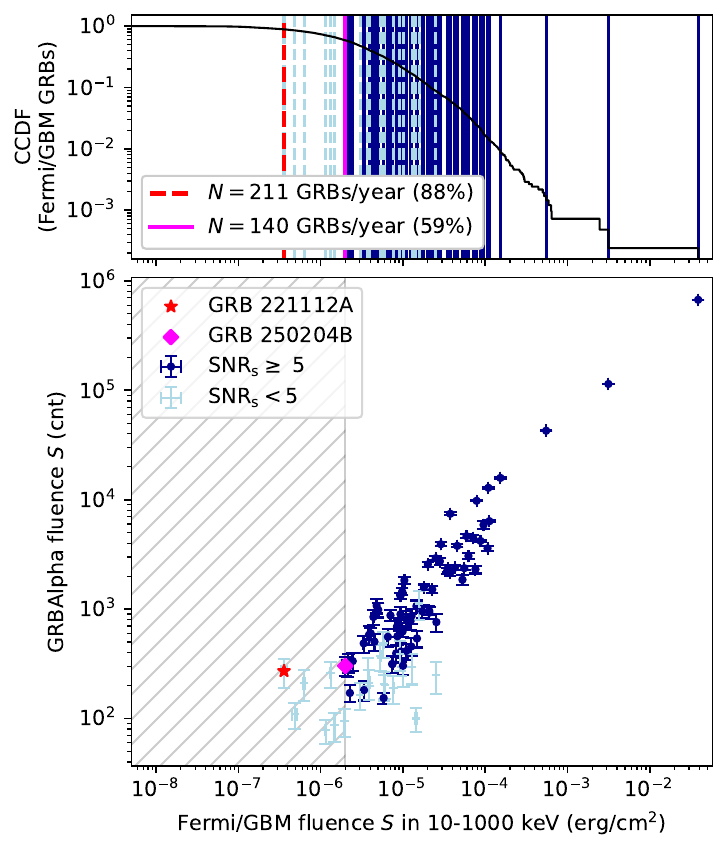}
    \end{minipage}
    \hfill
    \begin{minipage}[b]{0.49\linewidth}
        \centering
        \includegraphics[width=\textwidth]{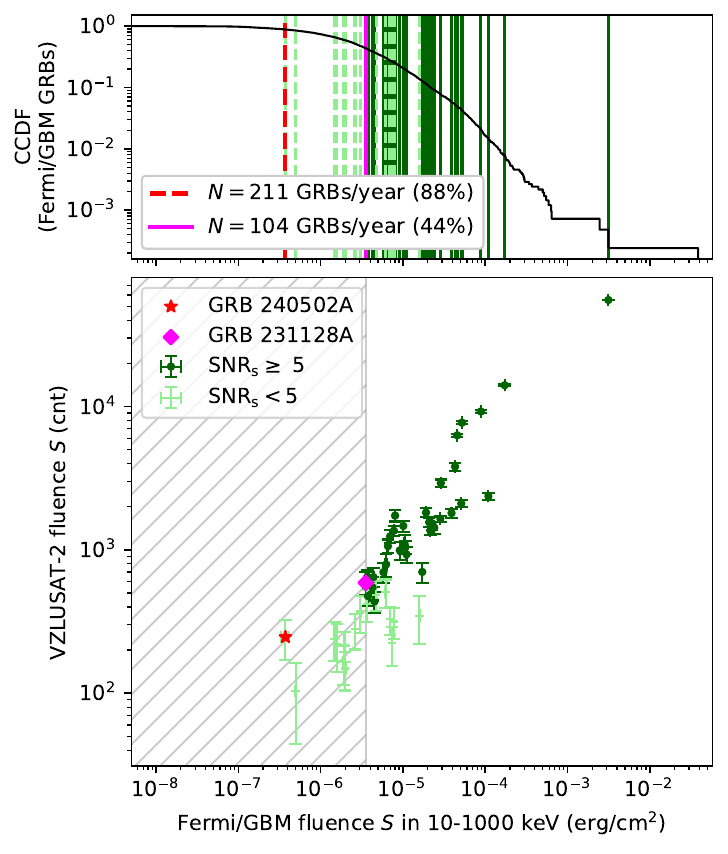}
    \end{minipage}
    \caption{
    Same as Fig.~\ref{fig:flux_gbm} but showing fluence $S$ instead of peak flux $P$.
    }
    \label{fig:fluence_gbm}
\end{figure*}

Comparison of the peak flux and fluence measured by the CubeSat missions and Fermi/GBM is depicted in Fig.~\ref{fig:flux_gbm} and \ref{fig:fluence_gbm}. Bottom panels compare the raw values obtained from GRBAlpha or VZLUSAT-2 with GBM while top panels place our detections in the context of all GRBs detected by GBM since launch of the Fermi satellite until 18 November 2025. 
Note that the peak flux and fluence were calculated in different energy range by each mission, 
the GBM peak flux is the physical photon flux and its fluence values are energy ﬂuence whereas for the CubeSats we report the detected flux and fluence in units of counts/s and counts, respectively. 
The FERMIGBRST catalog does not include the photon fluence and spectral parameters for 26~\% of GRBs from the subsamples of joint GBM and CubeSat GRBs, specifically, 24/90 GRBs from the GBM--GRBAlpha subsample and 14/52 GRBs from the GBM--VZLUSAT-2 subsample. In favor of a~larger sample size, we use the GBM energy fluence instead of photon fluence.
The positive correlations are rather dispersed due to the lack of attitude information for GRBAlpha and VZLUSAT-2, higher background around the Earth's radiation belts, and the difference in the energy bands.

GRBAlpha observed the lowest peak flux for GRB 221112A for which Fermi/GBM measured flux of $P=1.3\pm0.1\rm~ph~cm^{-2}~s^{-1}$. This belongs to the faintest $6\%$ of Fermi/GBM GRBs.
If we only consider GRBs detected with significance higher than $5\sigma$, the faintest GRB observed was 
GRB 230305A with Fermi/GBM $P=4.4\pm0.3\rm~ph~cm^{-2}~s^{-1}$ which falls close to the middle of the GBM peak flux distribution.
For VZLUSAT-2, the lowest recorded peak flux was for GRB 240502A with the Fermi/GBM measured flux $P=0.90\pm0.16\rm~ph~cm^{-2}~s^{-1}$. This is within $2\%$ of the faintest GBM GRBs.
Considering only $5\sigma$ detections, GRB 220912A is the weakest one with the Fermi/GBM $P=3.5\pm0.2\rm~ph~cm^{-2}~s^{-1}$, falling into the fainter half of GBM GRBs.

In terms of Fermi/GBM fluence, the faintest GRBAlpha sub-threshold detection (GRBAlpha $SNR<5\sigma$) is GRB 221112A with Fermi/GBM $S=(3.6\pm0.3)\cdot10^{-7}\rm~erg~cm^{-2}$
and GRB 250204B with Fermi/GBM $S=(2.00\pm0.03)\cdot10^{-6}\rm~erg~cm^{-2}$ from the significant subsample (GRBAlpha $SNR\geq5\sigma$). These events fall into the faintest $12\%$ and $41\%$ of all GBM GRBs, respectively.
VZLUSAT-2 observed the lowest fluence for GRB 240502A with the Fermi/GBM $S=(3.7\pm0.4)\cdot10^{-7}\rm~erg~cm^{-2}$ which belongs to the brightest $88\%$ of GBM GRBs. 
From the significant subsample, the faintest GRB is GRB 231128A with the Fermi/GBM $S=(3.53\pm0.06)\cdot10^{-6}\rm~erg~cm^{-2}$, falling into the brighter half of GBM GRBs.

The fact that some bright GRBs were detected as sub-threshold by the CubeSats is likely due to an off-axis orientation of the detector with respect to the source location. 
Among the sub-threshold detections from the peak flux point of view, some events were long and thus significant in terms of fluence.
Similarly, few short events were detected significantly at the peak but their SNR during the entire duration is lower than $5\sigma$. This happens especially for ``one-bin'' detections where the relative uncertainty of the fluence is larger due to background fluctuations (see Appendix~\ref{appendix:calculations}).
Nevertheless, these events would likely be triggered, just on a~different timescale.

Spectral properties of the faintest GRBs could potentially reveal information about what type of transients, if any, is the detector on board the CubeSats most sensitive to. 
To study this, 
we select those GRBs that were detected as sub-threshold both at the peak and during their entire duration. From these we further select the ones with GBM peak flux and fluence lower than that of the faintest significant GRB. That is, we select GRBs that lie in the gray hatched part of both Fig.~\ref{fig:flux_gbm} and \ref{fig:fluence_gbm}, separately for GRBAlpha and VZLUSAT-2.
Four GRBAlpha and three VZLUSAT-2 GRBs fall into this group. Their spectral parameters, reported by the Fermi/GBM team, are summarized in Tab.~\ref{tab:faint_grbs} along with median values from Fermi/GBM 10-year spectral catalog \citep{GBM_spec_cat}. 
From the long GRBs, three out of six differ from the median outside the $1\sigma$ confidence intervals. 
GRB 221112A was best fit by a power law model (PLAW) and its spectrum is slightly harder than a~median power law spectrum. 
GRB 240502A and GRB 240814B were best fit by a~power law model with an exponential cutoff (also called Comptonized model, COMP). 
GRB 240502A has the power law index $\alpha$ higher than that of the median GRB while the peak energy $E_{\rm p}$ is consistent with the median GBM GRB.  
For GRB 240814B, $\alpha$ is higher and $E_{\rm p}$ is lower than the median values. 
The short GRB 221120A is consistent with a median short GBM GRB.
This result does not point to any specific type of events the detector would be more sensitive to. 
Moreover, the sample size is very small, so we cannot draw any firm conclusions.

\begin{deluxetable*}{ccccccc}\label{tab:faint_grbs}
\tablecaption{Spectral index $\alpha$ and peak energy $E_{\rm p}$ of the faintest GRBs observed by GRBAlpha and VZLUSAT-2 and their comparison with the median from Fermi/GBM observations.}
\tablehead{
\colhead{event} & \colhead{type} & \colhead{mission} & \colhead{best fit model} & \colhead{$\alpha$} & \colhead{$E_{\rm p}$ (keV)} & \colhead{reference}
}
\startdata
GRB~221112A & long & GRBAlpha & PLAW & $-1.33\pm0.06$ & -- & FERMIGBRST \\ 
GRB 221120A & short & VZLUSAT-2 & COMP & $-0.07\pm0.4$ & $530\pm90$ & FERMIGBRST \\ 
GRB 240222A & long & GRBAlpha & PLAW & $-1.4\pm0.1$ & -- & \cite{GRB240222A_spectrum} \\ 
GRB 240502A & long & VZLUSAT-2 & COMP & $1.56\pm1.10$ & $151\pm21$ & \cite{GRB240502A_spectrum} \\   
GRB~240814B & long & GRBAlpha & COMP & $-0.33\pm0.05$ & $70\pm4$ & FERMIGBRST \\ 
GRB 241212A & long & VZLUSAT-2 & COMP & $-1.1\pm0.1$ & $305\pm59$ & \cite{GRB241212A_spectrum} \\ 
GRB~250109A & long & GRBAlpha & PLAW & $-1.57\pm0.05$ & -- & \cite{GRB250109A_spectrum} \\ median GRB & long & Fermi/GBM & PLAW & $-1.58_{-0.18}^{+0.15}$ & -- & \cite{GBM_spec_cat} \\       
median GRB & long & Fermi/GBM & COMP & $-1.01_{-0.35}^{+0.39}$ & $205_{-109}^{+374} $ & \cite{GBM_spec_cat}  \\       
median GRB & short & Fermi/GBM & COMP & $-0.39_{-0.46}^{+0.63}$ & $532_{-316}^{+732}$ & \cite{GBM_spec_cat} \\       
\enddata
\end{deluxetable*}

\subsection{Non-detections} \label{sec:nondets}

\begin{figure*}
    \hfill
    \begin{minipage}[b]{0.49\linewidth}
        \centering
        \includegraphics[width=\linewidth]{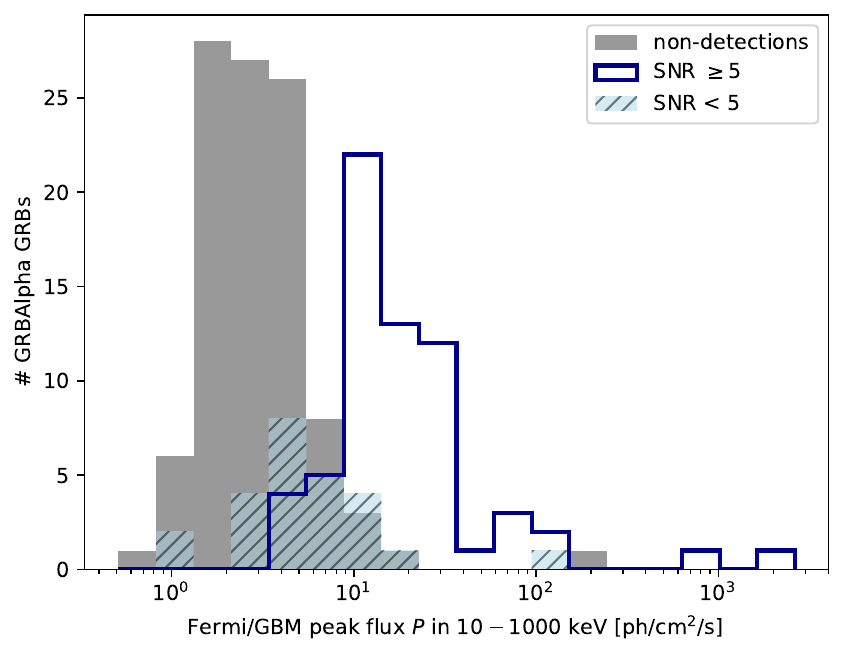}
    \end{minipage}
    \hfill
    \begin{minipage}[b]{0.49\linewidth}
        \centering
        \includegraphics[width=\textwidth]{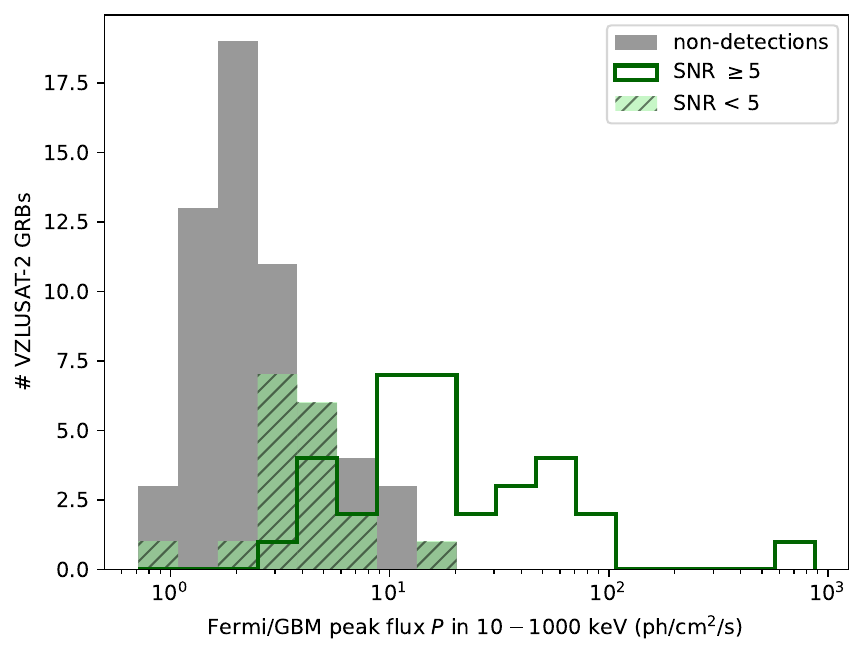}
    \end{minipage}
    \caption{
    Distributions of the peak flux $P$ of GRBAlpha (left) and VZLUSAT-2 (right) GRBs (in blue and green) compared to the ones which were not detected although the CubeSat was in low background and the event was not occulted by the Earth (in gray).
    }
        \label{fig:nondetections_flux}
\end{figure*}

\begin{figure*}
    \hfill
    \begin{minipage}[b]{0.49\linewidth}
        \centering
        \includegraphics[width=\linewidth]{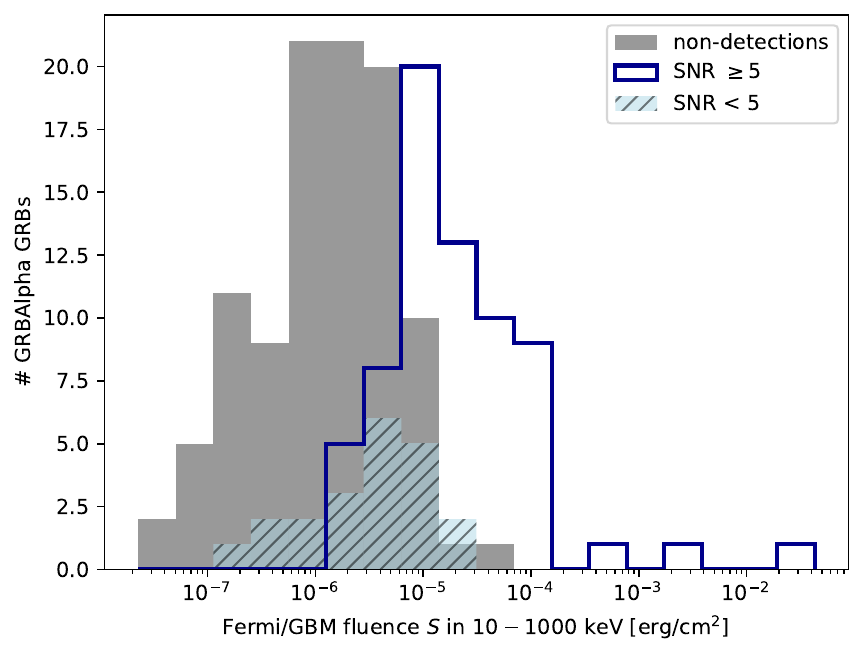}
    \end{minipage}
    \hfill
    \begin{minipage}[b]{0.49\linewidth}
        \centering
        \includegraphics[width=\textwidth]{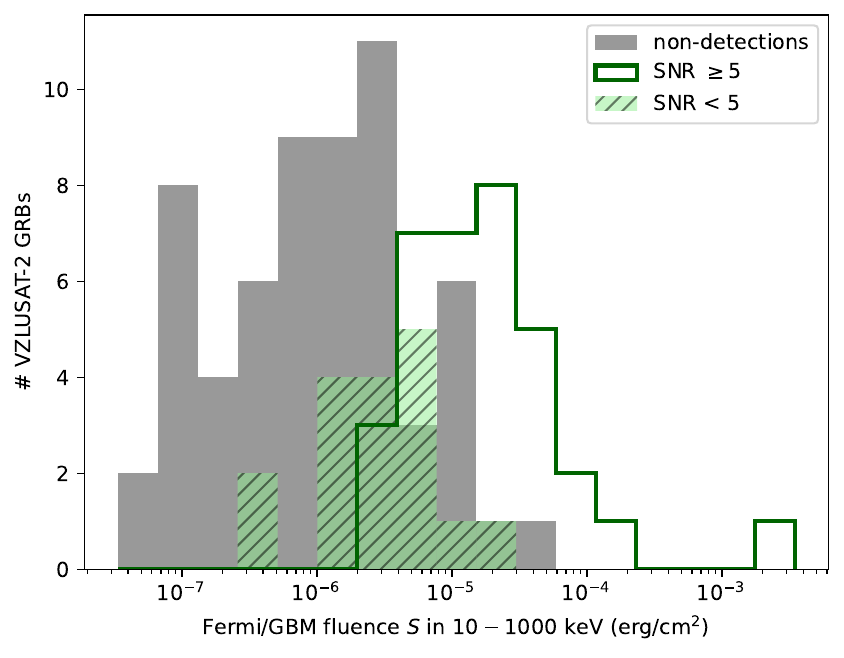}
    \end{minipage}
    \caption{
    Same as Fig.~\ref{fig:nondetections_flux} but showing fluence $S$ instead of peak flux $P$.
    }
        \label{fig:nondetections_fluence}
\end{figure*}

Besides the detections themselves, it is also interesting to look at why we did not see the rest of the GRBs.  
To study this, we created a sample of Fermi/GBM GRBs for which data acquired by GRBAlpha or VZLUSAT-2 are available. From the list we removed the ones that were detected by the CubeSats and those that were occulted by the Earth. For the remaining triggers in the sample we manually inspected the data  
and dropped those with high or variable background around the trigger time. The remaining sub-sample represents GRBs in the field of view of GRBAlpha or VZLUSAT-2 while the satellite was in low and stable background.
Fig.~\ref{fig:nondetections_flux} and \ref{fig:nondetections_fluence} compare the distributions of the peak flux and fluence of our sub-threshold and significant detections with the non-detected GRBs.  
Out of the GRBAlpha non-detections, $40~\%$ has peak flux or fluence higher than the GRB with the lowest GBM peak flux or fluence which was detected by GRBAlpha with $SNR\geq5\sigma$.
In case of VZLUSAT-2 it is $30~\%$. 

\section{Summary and Discussion} \label{sec:discussion_summary}

This paper presents gamma-ray transients observed by the 1U CubeSat GRBAlpha and the 3U CubeSat VZLUSAT-2. 
In total 173 GRBs, 164 solar flares, six bursts from soft gamma repeaters SGR 1806-20 and SGR J1935+2154 and one outburst from an X-ray binary system LS V +44 17 / RX J0440.9+4431 were observed. 
GRBAlpha and VZLUSAT-2 were the longest-lasting and most successful nanosatellite missions aiming to observe GRBs. They acquired the largest sample of gamma-ray transients observed by CubeSats with on average one GRB detected per week, at times less than one hour apart. Despite losing one of its two radio transceivers less than six months after the launch, GRBAlpha did regular measurements until less than three weeks before the re-entry. It detected the brightest-of-all-time GRB 221009A without saturation, contributed to IPN localization and both missions observed high-redshift GRBs up to $z=4.2$.  
Furthermore, 17 GRBs from this catalog were not observed by any mission providing localization on its own and the joint detection by GRBAlpha or VZLUSAT-2 is important for an~independent confirmation of their astrophysical origin.

When looking at the faintest detections, the empirical sensitivity in terms of what fraction of GBM GRBs could be observed by our detector exceeds $90\%$. While this is very impressive considering the much smaller effective area compared to that of GBM, the sub-threshold detections were confirmed by cross-correlation with the light curves observed by GBM and other missions. Due to their low significance, part of them would most likely not be triggered by an automatic algorithm on board. 
As GRBAlpha carried only one flat detector located at one side of the satellite, the detector sensitivity is highly dependent on the satellite's unknown attitude. Although VZLUSAT-2 carried two detector units, often only one of them was in operation.
Thus, the possible detection of over $90\%$ of GBM GRBs by one detector unit can be viewed as an optimistic estimate.

The most probable reason for the ``bright'' non-detections presented in Sec.~\ref{sec:nondets} is an off-axis orientation of the detector with respect to the source. The maximum effective area is 54~cm$^2$ for an on-axis direction but it drops below $\sim5\rm~cm^2$ for an edge-on direction. Moreover, the SiPMs are covered in lead shielding thus the effective area from one side is practically zero. This is likely what affected the detection of the bright GRB 241201A where GRBAlpha only observed the second fainter pulse and not the initial bright one although the background level was low and stable and the location region of the GRB was not occulted by the Earth. However, since we lack the attitude information we cannot confirm nor reject this hypothesis.

Satellites on a 500~km Sun-synchronous orbit spend around $40~\%$ of the time in high background regions due to the Earth's radiation belts \citep{Ripa2021}.
Assuming the instrument is omnidirectional, the field of view from 500~km is approximately 8.6~sr. Correcting the detection rate of one GRB per week 
for a~full sky coverage and a~100~\% duty cycle, we get $\sim125$ GRBs per year which would be detected by a constellation of GRBAlpha-like satellites providing an all-sky coverage from low background regions at all times. This is approximately half of GRBs observed by Fermi/GBM. 
If we only consider significant detections ($SNR\geq5\sigma$), GRBAlpha observed one GRB every nine days. This yields a detection rate of $\sim100$ GRBs per year.
Similarly, if we scale our observed detection rate of significant GRBs to the Fermi orbit (550~km and 85~\% duty cycle), we get 57 GRBs/year. 
However, the expected rate from the comparison with Fermi/GBM fluence shown in Fig.~\ref{fig:fluence_gbm} is more than double. Note, that this estimate of 140 GRBs/year assumes a~near-on-axis detection since GBM consists of 12 NaI detectors, each oriented in a~different direction such that the viewing angle of at least one detector is always below 60 degrees.
The estimate can be improved using the distribution of non-detections (Fig.~\ref{fig:nondetections_fluence}, left). 
We took all GBM GRBs with fluence higher that that of GRB 250204B (i.e., GRB with the lowest GBM-measured fluence from the sample of significant GRBs observed by GRBAlpha, see Sec.~\ref{sec:gbm_flux}).
Then we calculated the ratio of the fluence distribution of GRBAlpha significant detections to that of all GBM GRBs that GRBAlpha data are available for and that were not occulted by the Earth while the CubeSat was in low background region (i.e., ratio of the dark blue distribution shown in the left panel of Fig.~\ref{fig:nondetections_fluence} to the sum of all three distributions), again only considering GRBs with GBM-measured fluence higher than that of GRB 250204B.
Finally, we down-scaled the former distribution by the latter.
After this correction, the final estimate is 69 GRBs observed every year by a~GRBAlpha-like satellite on Fermi orbit. 
The remaining difference of 12 GRBs may be caused by various factors such as our insensitivity to low energies due to high thermal noise of SiPMs below $\sim100$~keV, the fact that GRBAlpha did not provide truly continuous measurements and that its detector is flat unlike GBM's thicker NaI detectors. 

\cite{CAMELOT} showed that a constellation of nine 3U CubeSats with detectors  
fully occupying two perpendicular sides would allow simultaneous coverage of 97.8~\% of the sky by at least 3 satellites on SSO at an altitude of 550 km. Assuming a 60~\% duty cycle due to high background in polar regions and SAA, at least 15 satellites would be needed to compensate for the loss of observing time.
We note that although the radiation belts significantly increase the measured background and thus decrease the chance of detecting typical GRBs, it is still beneficial to do observations in these regions. Many GRBs observed by GRBAlpha and VZLUSAT-2 were detected close to these regions, including the two brightest ones ever observed.

The original design of the CAMELOT constellation consists of 3U CubeSats with four large detectors, each composed of a CsI(Tl) scintillator with a size of 15 $\times$ 7.5 $\times$ 0.5~cm$^3$, placed at two perpendicular sides of the satellite body. \cite{CAMELOT_effective_area} simulated the maximum effective area of one such satellite to be $\sim340$~cm$^2$. This is larger than $\sim100$~cm$^2$ of one Fermi/GBM NaI detector \citep{FermiGBM} and comparable to its average over the unocculted sky of $\sim400$~cm$^2$ \citep{Stamatikos2009}. 
This indicates that a constellation of such nanosatellites would reach, if not exceed, the sensitivity of Fermi/GBM.

This catalog demonstrates the capabilities of small and quickly developed satellite missions with latest technologies for gamma-ray observations and their importance for multi-messenger astrophysics.
We show that tiny, cheap, and quickly developed satellites can contribute to larger missions and open a new pathway for future nanosatellite constellations monitoring the gamma-ray sky.

\begin{acknowledgments}
This work was supported by the Czech Science Foundation (GAČR) project No. 24-11487J and by the Internal Grant Agency of the Brno University of Technology under project No. FEKT-S-23-8191. MD is a Brno Ph.D. Talent Scholarship Holder---Funded by the Brno City Municipality.
\end{acknowledgments}

\break
\appendix

\section{Calculation of Observational Properties} \label{appendix:calculations}

Prior to any analysis, the raw data is normalized to count rate $CR$ as 
\begin{equation}
    CR = \frac{C}{\Delta t},
    \label{eq:count_rate}
\end{equation}
where $C$ is the number of counts detected during the resolution time $\Delta t$ which is the exposure time of each observation. 
Assuming that the detected counts follow the Poisson distribution, the $1\sigma$ uncertainty of the count rate is
\begin{equation}
    \sigma_{CR} = \frac{\sqrt{C}}{\Delta t} = \sqrt{\frac{CR}{\Delta t}}.
    \label{eq:count_rate_error}
\end{equation}

\subsection{Background Fitting}

The background is modeled separately for each energy band defined in Tab.~\ref{tab:energy_bands} and Fig.~\ref{fig:ebands-change-in-time}, as well as in the full combined band.
For each burst, we manually determine its start $T_{100,\rm s}$ and end time $T_{100,\rm e}$, and typically two background intervals; one before and one after the event. In few cases of very long emission episodes we included a third background interval between individual pulses to constrain the background model. These intervals are then fitted by polynomial functions of the first, second and third order. Due to a satellite rotation, usually between 5--20 deg/s, we sometimes observe periodic patterns in GRBAlpha's background. In such cases we add a cosine function to the polynomials. 
The fits are done by minimizing $\chi^2$ and the best-fit model is selected as the one with the lowest reduced $\chi^2$.
This best-fit background model is then subtracted from all data points and background-subtracted light curve is produced. In further analysis, we work with these source light curves.

\subsection{Duration}
\label{sec:t90}

We estimate the duration of all types of transients in terms of the $T_{50}$ and $T_{90}$ durations proposed for GRBs by \cite{Kouveliotou93}. 
Since most transients are only observed in the energy bands 0 and 1 (Tab.~\ref{tab:energy_bands}), we use this energy range to calculate the durations.

The observed duration of short events is restricted by the temporal resolution of the measurement while the duration uncertainty of longer bursts is given predominantly by statistical fluctuations. The uncertainty $\sigma_{T_{90}}$ (and analogically $\sigma_{T_{50}}$) is therefore calculated as a superposition of the two:
\begin{equation}
    \sigma_{T_{90}} = \sqrt{\sigma_{stat}^2 + (\Delta t)^2}.
\end{equation}

\noindent
We estimate the statistical uncertainty $\sigma_{stat}$ via a Monte Carlo simulation. For every event, we reproduce 10~000 simulated raw light curves, i.e., before background subtraction, with added Poisson noise to each temporal bin of the actual measured light curve. For each of these simulated light curves we then calculate the $T_{90}$ and $T_{50}$ while the background intervals and the $T_{100}$ of an event remain fixed. The median of the distribution is then taken as the final $T_{90}$ ($T_{50}$) and the lower and upper $1\sigma$ uncertainties are computed as the 0.159-th and 0.841-th quantiles, respectively.

\subsection{Peak Flux and Fluence}

We define the peak flux $P$ of a burst as the number of detected counts per second at a~peak time determined from the entire energy range covering all energy bands. The exact value and its uncertainty are calculated according to Eq.~\ref{eq:count_rate} and \ref{eq:count_rate_error}. 

The fluence $S$ is calculated from a cumulative background-subtracted light curve as a difference of means during the post-burst and pre-burst background regions \citep{Koshut96}. Its uncertainty is given by two components; statistical uncertainty and background fluctuations. The background uncertainty is a superposition of standard deviations of the pre- and post-burst background regions. The statistical uncertainty is given by the Poisson statistics as

\begin{equation}
    \sigma^2_{stat}=\sum_{i=t_s}^{t_f} C_i,
\end{equation}
where $i$ goes through all temporal bins from the time the cumulative counts reach the mean value of the pre-burst background region for the first time to the time when it reaches the post-burst value. The final fluence uncertainty is then calculated as a superposition of the two;

\begin{equation}
    \sigma_S=\sqrt{\sigma_{stat}^2+\sigma_{bg}^2}.
\end{equation}

\subsection{Hardness Ratio}
\label{sec:hardness_ratio}

The hardness ratio $HR$ is defined as the ratio of fluence in harder and softer energy band. Because most of the transients were detected only in the two lowest energy bands, we only include the $HR$ between these two bands, i.e.
\begin{equation}
    HR = \frac{S_1}{S_0}.
\end{equation}
Its uncertainty is computed according to the error propagation theory as
\begin{equation}
    \sigma_{HR} = HR\sqrt{\left(\frac{\sigma_{S_1}}{S_1}\right)^2 + \left(\frac{\sigma_{S_0}}{S_0}\right)^2} .
\end{equation}

\subsection{Detection Significance}

We calculate the significance of a detection in terms of the signal-to-noise ratio $SNR_p$ at the peak time and $SNR_s$ within the entire duration of a burst as the peak count rate or fluence divided by its uncertainty, i.e.,
\begin{equation}
    SNR_p = \frac{P}{\sigma_P}, \hspace{0.5cm} SNR_s = \frac{S}{\sigma_S}.
    \label{eq:snr}
\end{equation}

\bibliography{sample631}{}
\bibliographystyle{aasjournal}

\end{document}